# Cosmic-Ray-driven Multiphase Gas Formed via Thermal Instability

Xiaoshan Huang (黄小珊)[1,2] , Yan-fei Jiang (姜燕飞)[2] , and Shane W. Davis[1]
[1] Department of Astronomy, University of Virginia, Charlottesville, VA 22904, USA; xh2pm@virginia.edu
[2] Center for Computational Astrophysics, Flatiron Institute, 162 Fifth Avenue, New York, NY 10010, USA

## Abstract

Cosmic rays (CRs) are an important energy source in the circumgalactic medium that impact the multiphase gas structure and dynamics. We perform two-dimensional CR-magnetohydrodynamic simulations to investigate the role of CRs in accelerating multiphase gas formed via thermal instability. We compare outflows driven by CRs to those driven by a hot wind with equivalent momentum. We find that CR-driven outflow produces lower density contrast between cold and hot gas due to nonthermal pressure support, and yields a more filamentary cloud morphology. While entrainment in a hot wind can lead to cold gas increasing due to efficient cooling, CRs tend to suppress cold gas growth. The mechanism of this suppression depends on magnetic field strength, with CRs either reducing cooling or shredding the clouds by differential acceleration. Despite the suppression of cold gas growth, CRs are able to launch the cold clouds to observed velocities without rapid destruction. The dynamical interaction between CRs and multiphase gas is also sensitive to the magnetic field strength. In relatively strong fields, the CRs are more important for direct momentum input to cold gas. In relatively weak fields, the CRs impact gas primarily by heating, which modifies gas pressure.

*Unified Astronomy Thesaurus concepts:* Cosmic rays (329); Circumgalactic medium (1879); Interstellar medium (847); Galactic winds (572)

## 1. Introduction

The circumgalactic medium (CGM) is the halo of gas lying outside the galactic disk but within the virial radius. The CGM is an important baryon component of the galaxy, playing a key role in gas cycling by interactions with gas inflow and outflow, and affecting star formation fueling and feedback (Veilleux et al. 2005; Tumlinson et al. 2017). Tracing the structure and dynamics of CGM gas is important for us to understand galaxy evolution. One of the main techniques probing CGM composition and kinematics is through absorption line spectroscopy (Tumlinson et al. 2017). The hot $\sim 10^6$ K CGM gas has long been known as the "galactic corona." Recent observations suggest the existence of $\sim 10^5$ K intermediate-temperature gas (Wakker et al. 2012) that can reach $\sim 100\,{\rm km\,s^{-1}}$ bulk velocity. On the lower temperature end, the cold gas traced by low ionization lines is likely to be clumpy, while warm gas traced by high ionization lines is diffuse and potentially in coherent large structures up to kiloparsec scales (Bish et al. 2019; Werk et al. 2019). The COS-Halo survey reveals the presence of multiphase gas in other galaxies (Tumlinson et al. 2013; Werk et al. 2014). Observations seem to suggest the picture of outflow including fast-moving cold $\sim 10^4$ K clouds embedded in hot $\sim 10^6$ K background gas, with intermediate-temperature gas around $10^5$ K. Understanding this multiphase outflow is essential to CGM physics and galaxy evolution.

The origin and dynamics of cold CGM gas is an important puzzle. The cold clumps can either be launched from the gas disk (Klein et al. 1994; Cooper et al. 2009; Scannapieco & Brüggen 2015; McCourt et al. 2015; Gronke & Oh 2018; Scannapieco et al. 2020) or formed in situ from the hot gas via thermal instability (Field 1965; Sánchez-Salcedo et al. 2002; Sharma et al. 2010b; McCourt et al. 2012; Pal Choudhury et al. 2019; Girichidis et al. 2021). In a realistic CGM environment, it is likely that both mechanisms are operating. While there are extensive efforts to disentangle cloud launching and cold gas formation, reconstructing the interplay between the two processes is also important to bridge cloud-scale physics with a broader picture of multiphase outflow in a dynamical CGM environment.

Cosmic rays (CRs) are also a potentially important energy source in the CGM, which impacts both the cold cloud acceleration and thermal instability. These charged particles originate from supernovae explosions and propagate through the CGM, imparting energy and momentum to the gas. Recent numerical and theoretical work suggests that CRs can alter the pressure balance and phase structure in the CGM (Salem et al. 2016; Ji et al. 2020; Hopkins et al. 2020), and launch outflow or potentially prevent accreting inflows (Booth et al. 2013; Ruszkowski et al. 2017; Crocker et al. 2021b; Quataert et al. 2022; Hopkins et al. 2021). CRs may modulate thermal instability (Sharma et al. 2010b; Butsky et al. 2020), launch cold clouds (Wiener et al. 2019; Brüggen & Scannapieco 2020; Bustard & Zweibel 2021), and produce potentially observable modifications to absorption line profiles (Butsky et al. 2021).

Although CRs are an interesting and important component in galactic evolution, incorporating CR physics in numerical simulations is not trivial. This is partly due to our limited knowledge about CR transport itself. Different CR transport models are assumed such as self-confinement streaming (Wentzel 1974; Crocker et al. 2021a; Quataert et al. 2022), super-Alfvénic streaming associated with turbulent damping (Ruszkowski et al. 2017; Holguin et al. 2019; Hopkins et al. 2020), and isotropic or anisotropic diffusion (Salem et al. 2016; Wiener et al. 2017). A number of works suggest that CR transport models significantly impact CR interaction with gas and its ability to drive galactic outflow (Uhlig et al. 2012;

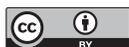

Butsky & Quinn 2018; Buck et al. 2020; Huang & Davis 2022).

At the same time, the numerical difficulties associated with the moments equations make the implementation of CR transport challenging. For example, Sharma et al. (2010a) adopted a regularization scheme to smooth out the step-function-like streaming velocity near CR pressure maximum. Jiang & Oh (2018) resolved this issue by solving two-moment formulation of the CR equations, facilitating the integration of CR transport in both streaming and diffusion limits. Thomas & Pfrommer (2019) and Chan et al. (2019) also explored CR two-moment implementations with slightly different approaches. Thomas & Pfrommer (2022) compared closure relations for CR two-moment equations and found that the impact is small in the streaming limit, where scattering is neglected. Hopkins et al. (2022) derived and implemented moment equations that can also handle CRs in the strong scattering limit with generalized source terms.

In this work, we primarily focus on CR transport in the self-confined streaming limit, with one supplemental CR diffusion run as comparison to the relatively fast streaming. We use the Athena++ implementation of Jiang & Oh (2018). By solving both CR energy and flux equations, the scheme is less diffusive than the single-moment method and better preserves the CR energy behavior where it decouples with gas.

Recent simulations provide us a physical picture of how CR streaming can accelerate cold gas via a "bottleneck" (Wiener et al. 2019; Brüggen & Scannapieco 2020; Bustard & Zweibel 2021; Huang & Davis 2022). By studying cloud–CR interaction in a relatively well-controlled environment, the cloud-scale simulations provide useful theoretical insights about CR acceleration, cloud-crushing timescales, and CR-modified density contrast in multiphase gas. They also raise concerns, however, due to the idealized setup. Particularly for CR streaming, it seems important to recover some key environmental factors in the CGM such as clumpiness, irregular cloud morphology, and nonuniform magnetic fields.

In addition to the dynamics, CR pressure may also impact the formation of multiphase gas. Studies show that unlike the classical isobaric thermal instability, CR mediated thermal instability can be an isochorical process (Sharma et al. 2010a; Kempski & Quataert 2020). The CR pressure provides nonthermal support to the cooling gas and prevents rapid contraction. Butsky et al. (2020) estimated that density contrast of the multiphase gas formed from thermal instability will be reduced due to the presence of CRs. Their work also suggests that in a CR-pressure-dominated halo, cold clouds are potentially an order of magnitude larger than what is predicted by the cloud scale in a purely thermal medium. Kempski & Quataert (2020) studied the effect of CRs on linear thermal instability, and found that ratio of CR to gas pressure controls the transition from isobaric to isochoric. Although CR heating does not directly change the instability growth rate in isobaric or isochoric limits, the perturbed CR heating make gas entropy mode oscillatory.

Given various complexities brought by CRs, in this work, we do not set the cold and hot gas in pressure balance; instead, we study the interaction of CRs with multiphase gas structure that formed in situ due to thermal instability. Thermal instability spontaneously yields $\sim 10^4$ K cold clouds in diffuse $\sim 10^6$ K hot gas with $\sim 10^5$ K interface, emulating the observed multiphase CGM gas (Sharma et al. 2010b; Jennings & Li 2021). A key focus is to study if CRs can accelerate cold gas without successively destroying it, especially in a more realistic environment where the background is not initially uniform. Another goal is to compare CR with a hot wind as outflow-driving mechanisms. While both mechanisms are thought to originate from supernovae, their interactions with multiphase gas can be intrinsically different, potentially leading to different outflow properties.

We introduce the simulation setup and relevant scaling in Section 2. We present the results in Section 3, where we compare CRs and a hot wind accelerating and modifying multiphase gas. In Section 4 we discuss CR streaming in turbulent magnetic fields, describe factors affecting cold gas survival, and connect our results with previous work. We summarize our conclusions in Section 5.

## 2. Simulation Setup

### 2.1. Equations

We solve the following equations in Athena++ with the cosmic-ray transfer module (Jiang & Oh 2018):

$$\frac{\partial \rho}{\partial t} + \nabla \cdot (\rho \mathbf{v}) = 0, \quad (1)$$

$$\frac{\partial (\rho \mathbf{v})}{\partial t} + \nabla \cdot (\rho \mathbf{v}\mathbf{v} - \mathbf{B}\mathbf{B} + \mathsf{P}^*)$$
$$= \sigma_{\rm CR} \cdot [\mathbf{F}_{\rm CR} - \mathbf{v} \cdot (E_{\rm CR}\mathsf{I} + \mathsf{P}_{\rm CR})], \quad (2)$$

$$\frac{\partial E}{\partial t} + \nabla \cdot [(E + P^*)\mathbf{v} - \mathbf{B}(\mathbf{B} \cdot \mathbf{v})] + \nabla \cdot \mathbf{Q}$$
$$= (\mathbf{v} + \mathbf{v}_{\rm s}) \cdot \sigma_{\rm CR} \cdot [\mathbf{F}_{\rm CR} - \mathbf{v} \cdot (E_{\rm CR}\mathsf{I} + \mathsf{P}_{\rm CR})] + Q_{\rm cool}, \quad (3)$$

$$\frac{\partial \mathbf{B}}{\partial t} - \nabla \times (\mathbf{v} \times \mathbf{B}) = 0, \quad (4)$$

$$\frac{\partial E_{\rm CR}}{\partial t} + \nabla \cdot \mathbf{F}_{\rm CR}$$
$$= -(\mathbf{v} + \mathbf{v}_{\rm s}) \cdot \sigma_{\rm CR} \cdot [\mathbf{F}_{\rm CR} - \mathbf{v} \cdot (E_{\rm CR}\mathsf{I} + \mathsf{P}_{\rm CR})] \quad (5)$$

$$\frac{1}{V_{\rm m}^2} \frac{\partial \mathbf{F}_{\rm CR}}{\partial t} + \nabla \cdot \mathsf{P}_{\rm CR}$$
$$= -\sigma_{\rm CR} \cdot [\mathbf{F}_{\rm CR} - \mathbf{v} \cdot (E_{\rm CR}\mathsf{I} + \mathsf{P}_{\rm CR})]. \quad (6)$$

Here $\rho$, $\mathbf{v}$, and $E$ are fluid density, velocity, and total energy, and $\mathbf{B}$ is magnetic field strength. $P^*$ is the sum of gas pressure and magnetic pressure, $\mathsf{P}^*$ is the corresponding pressure tensor. $\nabla \cdot \mathbf{Q}$ represents the energy change due to thermal conduction, where $\mathbf{Q} = -\kappa \nabla T$ is the heat flux and $\kappa$ is the thermal conductivity. In this work, the conductivity parallel and perpendicular to the magnetic field are $\kappa_\parallel = 2.0 \times 10^9$ erg K$^{-1}$ s$^{-1}$ cm$^{-1}$ and $\kappa_\perp = 4.0 \times 10^8$ erg K$^{-1}$ s$^{-1}$ cm$^{-1}$, respectively. Note that the $\kappa_\parallel/\kappa_\perp$ assumed here is potentially lower than the realistic value. We choose this ratio in order to resolve the Field length (Equation (8)) in both the $x$- and $y$-directions (Koyama & Inutsuka 2004).

Equations (5) and (6) are the CR momentum and energy equation, i.e., the first and second moment integration over the energy and angle of the CR advection–diffusion equation (Skilling 1971). The CR streaming velocity is $\mathbf{v}_{\rm s} = -\mathrm{sgn}(\mathbf{B} \cdot \nabla P_{\rm CR}) \mathbf{v}_A$. It has the magnitude of the Alfvén velocity and points opposite to the CR pressure gradient.

This CR transport description is based on the "self-confinement" assumption. In this framework, CRs excite Alfvén waves as they stream through the plasma. When the streaming velocity exceeds the local Alfvén velocity, the Alfvén wave is amplified and eventually saturates. As the Alfvén wave grows and perturbs the magnetic field, magnetic field irregularities will scatter the CRs and change their pitch angles, thus reducing the CR streaming velocity to near the Alfvén speed. Hence, the CRs must stream at the Alfvén speed and down the CR pressure gradient. The resulting wave damping enables the momentum and energy exchange between CRs and gas.

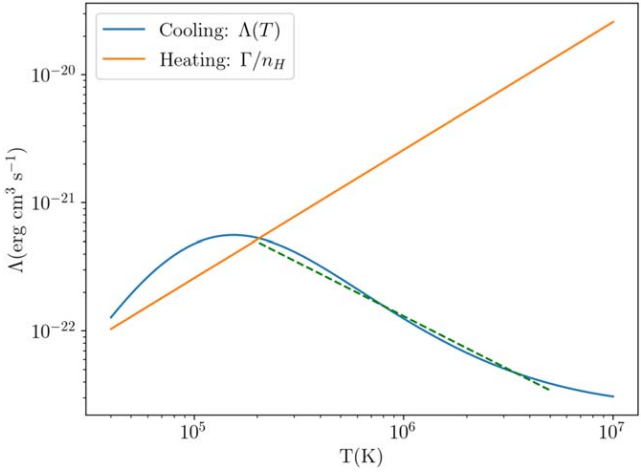

**Figure 1.** The cooling curve we adopted in all simulations from Wiener et al. (2019). We assume a constant pressure of $P_{\rm gas,0} = 2.33 \times 10^{-21}$ dyne cm$^{-2}$ in the plot. The blue solid line is the cooling term $\Lambda(T)$, while the orange solid line is the supplemental heating term $\Gamma/n_{\rm H}$. The green solid line is the power-law fitting to the unstable branch with index $-0.83$. We truncate the cooling and heating for gas with temperature $T < 4 \times 10^4$ K.

$E_{\rm CR}$ and $\boldsymbol{F}_{\rm CR}$ are the cosmic-ray energy density and flux. The two-moment scheme evolves the time-dependent Equation (6) to update $\boldsymbol{F}_{\rm CR}$, which differs from one-moment schemes that only evolve $E_{\rm CR}$ and prescribe a form for $\boldsymbol{F}_{\rm CR}$ based on $E_{\rm CR}$. Such treatment is less diffusive for calculating the streaming velocity near the CR pressure gradient maximum, where the discontinuous streaming velocity is better captured. $V_{\rm m}$ is the maximum CR propagation velocity, which is assumed to be constant. It replaces the speed of light $c$ in Equation (6) to relax the time step. As long as $V_{\rm m}$ is significantly larger than the maximum flow or Alfvén velocity, the impact to gas dynamics is limited.

$\sigma_{\rm CR}$ is the CR–gas interaction coefficient:

$$\sigma_{\rm CR}^{-1} = \sigma_{\rm CR}'^{-1} + \frac{\boldsymbol{B}}{|\boldsymbol{B} \cdot (\nabla \cdot \mathsf{P}_{\rm CR})|} \boldsymbol{v}_{\rm A} \cdot (E_{\rm CR}\mathsf{I} + \mathsf{P}_{\rm CR}), \qquad (7)$$

where $\sigma_{\rm CR}'^{-1} = \kappa_{\rm diff}$ is the conventional CR diffusion coefficient. The second term is an effective streaming coefficient $\sigma_{\rm str}^{-1}$. The format indicates that streaming requires a nonzero CR pressure gradient for the coupling between CR and gas.

$Q_{\rm cool}$ is the external energy source term. We assume optical thin cooling and supplemental heating is $-n_{\rm H}^2 \Lambda(T) + n_{\rm H}\Gamma$, where $n_{\rm H}$ is the gas number density. The cooling function $\Lambda(T)$ is an approximation to fit the CLOUDY data with solar metallicity from Wiener et al. (2019). The heating constant is initialized as $\Gamma = 6 \times 10^{-24}$ erg s$^{-1}$. We cut off the radiative cooling and supplemental heating for gas with temperature below $4 \times 10^4$ K. We plot the cooling function in Figure 1.

### 2.2. Characteristic Scales

Sharma et al. (2010b) studied the formation of cold filaments via thermal instability with adiabatic cosmic rays. They found that the fastest growing mode of thermal instability tends to be elongated along the magnetic field lines with a characteristic length scale of the Field length:

$$L_F = 2\pi \left[ \frac{\chi t_{\rm cool}}{d \ln(T^2/\Lambda)/d \ln T} \right]^{1/2}, \qquad (8)$$

where $\chi_\parallel = \kappa_\parallel / n_e k_{\rm B}$ ($\chi_\perp = \kappa_\perp / n_e k_{\rm B}$) is the scaled conductivity, and $t_{\rm cool}$ is the cooling timescale, and the denominator can be approximated by the power-law index of the unstable branch of the cooling function. In our simulations, the unstable branch corresponds to the temperature range $T \gtrsim 2 \times 10^5$ K with the negative slope in Figure 1, which can be approximated by a power law with an index of $-0.83$ (the green dashed line).

When a cloud is accelerated by a hot wind, an important timescale that roughly describes the time it takes the wind to deform the cold cloud is the cloud-crushing time (Klein et al. 1994):

$$t_{\rm cc} = \sqrt{\frac{\rho_{\rm c}}{\rho_{\rm h}}} \frac{R_{\rm c}}{v_{\rm hot}}, \qquad (9)$$

where $\rho_{\rm c}$ and $\rho_{\rm h}$ are the cold and hot gas density, respectively, and $v_{\rm hot}$ is the wind speed. For cold clouds formed in our simulation, the sizes of the cloud $R_{\rm c}$ are usually about one to a few $L_F$.

When the cloud is accelerated by CR streaming, we estimate the expected acceleration using the following simplified model. When the flow velocity is relatively small compared to the Alfvén velocity, the CR flux $F_{\rm CR} \approx 4 v_{\rm A} P_{\rm CR}$. If the time-dependent term in Equation (6) is relatively unimportant, the CR force on the gas is roughly the CR pressure gradient $\partial P_{\rm CR}/\partial x \approx (P_{\rm CR,l} - P_{\rm CR,r})/R_{\rm c}$. Assuming CR pressure on the irradiated side $P_{\rm CR,l}$ is significantly larger than the shaded side $P_{\rm CR,r}$, the CR acceleration can be approximated as:

$$a_{\rm CR} = f \frac{F_{\rm CR}}{4 v_{\rm A} L_F \rho_{\rm c}} \qquad (10)$$

where $f$ is a factor added to account for deviations from our assumptions.

### 2.3. Scaling and Initialization

In this work, we perform seven simulations in total. The simulations are labeled B2CR, B2CRdiff, B2CR_LC, B2HW, B1CR, B1HW, and B05CR. The naming convention uses "B2, B1, B05" to denote the initial magnetic field strength, while the following characters denote the acceleration mechanism, with "CR" for cosmic-ray streaming, "CRdiff" for cosmic-ray diffusion, and "HW" for hot wind. We also present an additional run B2CR_LC with slightly different conductivity than the other six simulations to explore the effect of varying the heating constant $\Gamma$ relative to other parameters.

We solve the dimensionless equations with the following scaling: the temperature unit is the initial equilibrium temperature $T_0 = 2.04 \times 10^5$ K. The density unit is $\rho_0$. The velocity unit is set to be the sound speed at $T_0$. The initial cooling time $t_{c,i}$ is:

$$t_{c,i} = \frac{k_B}{\mu m_p} \frac{\rho_0 T_0}{n_{H,0}\Gamma}. \quad (11)$$

The time unit is $t_0 = 10 t_{c,i}$, and the length unit $l_0 = 10 v_0 t_{c,i}$, where $k_B$ is the Boltzmann constant, $\mu = 0.6$ is the mean molecular weight, and $n_{H,0} = \rho_0/\mu m_p$ is the initial number density. The heating constant $\Gamma = n_{H,0}\Lambda(T_0)$.

We use the same initial equilibrium temperature $T_0 = 2.04 \times 10^5$ K for all simulations. However, we have a degree of freedom of rescaling the simulations to different density unit $\rho_0$. When rescaling $\rho_0$, we also change the corresponding heating constant $\Gamma$. We scale the cooling function to $\Lambda(T_0) = 5.28 \times 10^{-22}$ erg s$^{-1}$ cm$^3$, which is the cooling rate at $T_0$. We can write the dependence of $\rho_0$ and $\Gamma$, $l_0$ and $L_F$ as:

$$\Gamma = 5.27 \times 10^{-24} \text{ erg s}^{-1} \left[\frac{\Lambda(T)}{\Lambda(T_0)}\right]\left(\frac{\rho}{10^{-26} \text{ g cm}^{-3}}\right) \quad (12)$$

$$l_0 = 91.6 \text{ pc}\left(\frac{T}{T_0}\right)^{3/2}\left[\frac{\Lambda(T)}{\Lambda(T_0)}\right]^{-1}\left(\frac{\rho}{10^{-26} \text{ g cm}^{-3}}\right)^{-1} \quad (13)$$

$$L_F = 106.5 \text{ pc}\left(\frac{T}{T_0}\right)^{1/2}\left(\frac{\kappa_\parallel}{2 \times 10^9 \text{ erg K}^{-1} \text{s}^{-1} \text{cm}^{-1}}\right)^{1/2}$$
$$\times \left[\frac{\Lambda(T)}{\Lambda(T_0)}\right]^{-1/2}\left(\frac{\rho}{10^{-26} \text{ g cm}^{-3}}\right)^{-1}. \quad (14)$$

When reporting results in cgs units, we adopt $\rho_0 = 1.14 \times 10^{-26}$ g cm$^{-3}$, a length unit $l_0 = 2.48 \times 10^{20}$ cm, and a temperature unit $T_0 = 2.04 \times 10^5$ K for most simulations except for B2CR_LC. This implies a heating constant $\Gamma = 6 \times 10^{-24}$ erg s$^{-1}$. The velocity unit is set to be the sound speed at $T_0$, $v_0 = 5.29 \times 10^6$ cm s$^{-1}$, and the time unit corresponds to the sound crossing time $t_0 = 1.49$ Myr.

In B2CR_LC, we scale the simulation to $\rho_0 = 1.14 \times 10^{-27}$ g cm$^{-3}$, $l_0 = 2.48 \times 10^{21}$ cm, $T_0 = 2.04 \times 10^5$ K, and $\Gamma = 6 \times 10^{-25}$ erg s$^{-1}$. We lower $\kappa_\parallel = 2.0 \times 10^7$ erg K$^{-1}$ s$^{-1}$ cm$^{-1}$ and $\kappa_\perp = 4.0 \times 10^6$ erg K$^{-1}$ s$^{-1}$ cm$^{-1}$ in order to keep $L_F$ fixed and form clouds with similar physical sizes as other simulations (see Section 2.2). In the following sections, we report quantities in dimensionless code units with a prime, unless otherwise specified.

All simulations are two-dimensional, and the domain size is $L_x \times L_y = 400 l_0 \times 50 l_0$ with $8000 \times 1000$ cells on each side, given the equivalent resolution of 4.02 pc in both the $x$- and $y$-directions. In B2CR_LC, we use a smaller domain of $L_x \times L_y = 40 l_0 \times 5 l_0$ in order to resolve $L_F$ with the same resolution.

There are two main stages in each simulation: (1) the multiphase gas formation via thermal instability and (2) the acceleration by CR or hot wind. To model the multiphase gas formed in situ, we first perturb the initially uniform gas from thermal equilibrium, so that thermal instability generates multitemperature gas that remains roughly in pressure equilibrium. We assume a low background CR energy density in this stage. Once the multiphase gas reaches an approximate steady state, we inject a uniform CR flux or a hot wind from the left $x$-boundary to model the energy and momentum input from the star-forming region. Here we describe the two stages accordingly.

Initially, we place a uniform slab with density $\rho'_{\text{init}} = 1.0$ between $50 < x' < 150$, and set the temperature to be the equilibrium temperature $T_{\text{init}} = T_0$. The background gas density is set to a density floor $\rho_{\text{bkgd}} = 10^{-6}\rho_0$ and in pressure equilibrium with the slab. We add small random perturbations to the gas density in order to perturb it from the equilibrium, $\rho' = 1 + 0.1\delta\rho'$, where $\delta\rho'$ is randomly distributed between $-1$ and 1. Initially, the magnetic field is uniform along the $x$-direction with $\beta_{\text{init}} = 0.5$, 2.0, and 8.0 in simulations noted by B2, B1, and B05. We set a low background CR energy density $E'_{\text{CR,init}} = 0.3$.

The $y$-direction boundary conditions are periodic. In the $x$-direction, the boundary conditions are different in the multiphase gas formation stage and acceleration stage. In the first stage, the $x$-direction hydrodynamic and CR boundary conditions are outflow, where the ghost zones copy values from the last active zone. The only exception is B2CRdiff, where we set the velocity to be a single-direction outflow to avoid any unwanted diffusive flux from boundary. The ghost zones copy velocity from last active zone but set any inward velocity to be zero.

In the second stage, in B2CR, B2CRdiff, B1CR, and B05CR, we set the left hydro boundary to be reflecting. The left CR boundary fixes CR flux $F'_{\text{CR}}$ in the $x$-direction, and copies other variables from the last active zone. We set $F'_{\text{CR}} = 60$, corresponding to $F_{\text{CR}} \approx 2.5 \times 10^5 L_\odot \text{ kpc}^{-2}$. Using the CR flux estimate from Crocker et al. (2021a), this flux roughly corresponds to systems with surface star formation rate $\dot{\Sigma}_* \approx 0.05 M_\odot \text{ kpc}^{-2}\text{yr}^{-1}$. In B2HW, B1HW, we inject a hot wind at the left boundary with ghost zone density $\rho'_{\text{wind}}$ and velocity $v'_{\text{wind}}$. Other variables obey outflow boundary conditions (see the exact values in Section 3.2). The right hydrodynamic boundary is single-direction outflow that only allows outward flux. We also imposed a temperature upper limit of $T' < 200$ for gas in ghost zones to control the heat flux into the domain.

In both stages, the magnetic field boundary conditions are set to be continuous for the perpendicular component and zero across the boundary for the parallel component in the $x$-direction, and periodic in the $y$-direction. The thermal conduction boundaries are reflecting.

## 3. Results

We perform three main sets of simulations with differing initial magnetic field: strong field, $B'_{x,\text{init}} = 2$, $\beta_{\text{init}} = 0.5$ (B2CR and B2HW), $B'_{x,\text{init}} = 1$, $\beta_{\text{init}} = 2.0$ (B1HW and B1CR), and $B'_{x,\text{init}} = 0.5$, $\beta_{\text{init}} = 8.0$ (B05CR).

### 3.1. Formation of Multiphase Gas via Thermal Instability

The initial slab is in thermal equilibrium with small random perturbation. Thermal instability develops because we perturb the gas temperature to fall into the unstable branch of the cooling curve, so that when the gas cools and collapses, the cooling is enhanced and eventually triggers runaway cooling. With our initialization and choice of cooling function (see Section 2.3 and Figure 1), we estimate $L'_{F,\parallel} \approx 1.16$ in the $x$-direction and $L'_{F,\perp} \approx 0.38$ in the $y$-direction, which is well resolved by $\sim 24$ cells in the $x$-direction and $\sim 8$ cells in the $y$-direction.

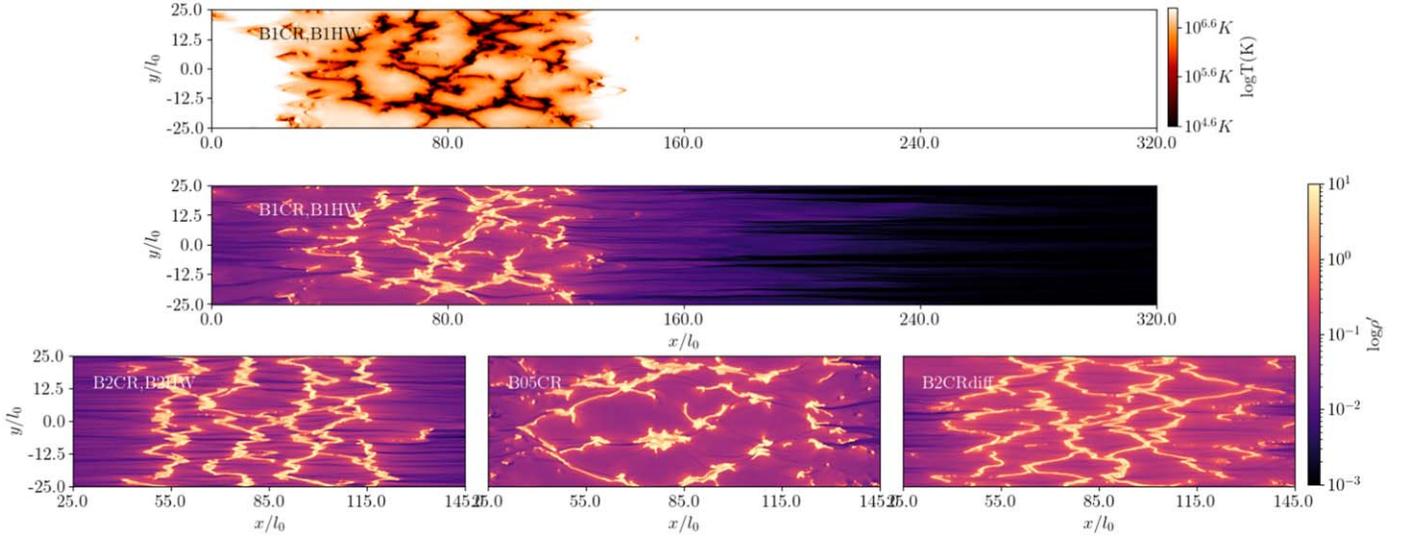

**Figure 2.** Multiphase gas formed via thermal instability. The first row shows a gas temperature snapshot for the fiducial runs B1CR and B1HW, while the second row shows the corresponding density snapshot at $t' = 30 \approx 300 t'_{c,i}$. The third row shows the snapshots before we inject CR flux or hot wind in other sets of simulations, from left to right: B2CR and B2HW at $t' = 30$, B05CR at $t' = 30$, B2CRdiff at $t' = 25$.

The initial cooling timescale for the perturbed gas is $t'_{c,i} \approx 0.1$. We run the simulation to $t' = t'_\parallel = 30 \approx 300 t'_{c,i}$, where $t'_\parallel$ denotes the time when we inject CR flux or hot wind. The gas evolution is in a quasi-steady state, and the gas density and temperature distribution are relatively constant. Figure 2 shows the gas density and temperature before we inject CRs or hot wind. In all simulations, the cold phase reaches the cooling function's lower temperature limit of $4 \times 10^4$ K. The hotter gas ($\sim 10^{6-7}$ K) expands and fills part of the domain but leaves $x' \gtrsim 200$ close to the initial density and temperature.

The cold gas is clumped and forms clouds with sizes of roughly a few Field lengths. The overdense structures (e.g., filaments) preferentially align with the magnetic field, and there is a trend for structure to become less oriented when the magnetic field is weaker. The overall morphology is consistent with previous MHD thermal instability studies (Sharma et al. 2010b; Jennings & Li 2021).

The density contrast is related to the balance between gas, CRs, and magnetic pressure. The temperature contrast is primarily set by the cooling curve. Hereafter we define *cold gas* as gas with temperature $T \leqslant 7 \times 10^4$ K, corresponding to the dense, clumped gas; *intermediate-temperature gas* as $7 \times 10^4$ K $< T \leqslant 2 \times 10^5$ K, roughly corresponding to the temperature where cooling is maximized; *warm gas* as $2 \times 10^5$ K $< T \leqslant 10^6$ K, where supplemental heating dominates cooling; and *hot* gas as $T \geqslant 10^6$ K, where density is relatively low, and both the cooling and heating are moderate.

We expect that the spontaneously formed multiphase gas provides a more realistic initialization to study CR- and hot-wind-driven outflow. The perturbed magnetic field and irregular cloud morphology in our initialization are also important when studying CR streaming in a CGM-like environment. Moreover, the nonuniform pressure distribution and the continuous gas temperature range better approximate the multiphase gas in the CGM, adding complexity to the classical cold gas survival and acceleration problem.

### 3.2. Acceleration by Hot Wind: B1HW

We first show the results from B1HW and B1CR, a set of simulations with the same fiducial magnetic field. We implement a hot wind boundary condition so that it provides a comparable momentum flux to the CR flux in B1CR. In CR runs, we inject uniform, constant CR flux $F'_{CR} = 60$ from the left $x$-boundary at $t' = 30$ ($t' = 25$ for B2CRdiff). In B1HW and B2HW, the density of hot wind is set to be the average gas density at $x' \lesssim 10$, and yields $\rho'_w = 0.015$ in B2HW and $\rho'_w = 0.069$ in B1HW.

The wind velocity $v'_{wind} = 7.83$ in B2HW and $v'_{wind} = 5.55$ in B1HW. We calculate the wind velocity by $\rho'_w v'^2_w = F'_{CR}/4v'_A$ according to the following estimation. The momentum flux carried by the hot wind is assumed to be equivalent to the momentum flux of CR, which leads to $\rho'_w v'^2_w = \int f'_{CR} dx'$, where $f'_{CR}$ is the CR force. With the steady-state assumption, $\int f'_{CR} dx' \sim \int \nabla P_{CR} dx \sim P'_{CR,left} - P'_{CR,right}$, $P'_{CR,left}$ and $P'_{CR,right}$ are the CR pressures at the left and right boundaries, respectively. At a relatively early time, $P'_{CR,right} \ll P'_{CR,left}$. In the streaming limit, the injected CR flux $F'_{CR} \approx 4v'_A P'_{CR}$, where $v'_A = B'_x/\sqrt{\rho'_h}$ is the local Alfvén velocity, give the above estimation of $v_w$. The injected winds are subsonic in both simulations.

Figure 3 shows the gas density snapshots from B1HW. When accelerated by a thermal wind, the clouds are compressed and some clouds merge together. During acceleration, the clouds are stretched in the direction of the wind, but some filaments from initial thermal instability maintain filamentary morphology. The clouds formed are nearly spherical, with a typical diameter similar to a few initial Field lengths. The field lines wrap around the spherical clouds and form hoops, increasing local field strength and preventing deformation.

Figure 4 shows the cold gas mass evolution and average velocity. The cold gas evolution is generally consistent with quasi-linear acceleration with slight fluctuations. The cold gas

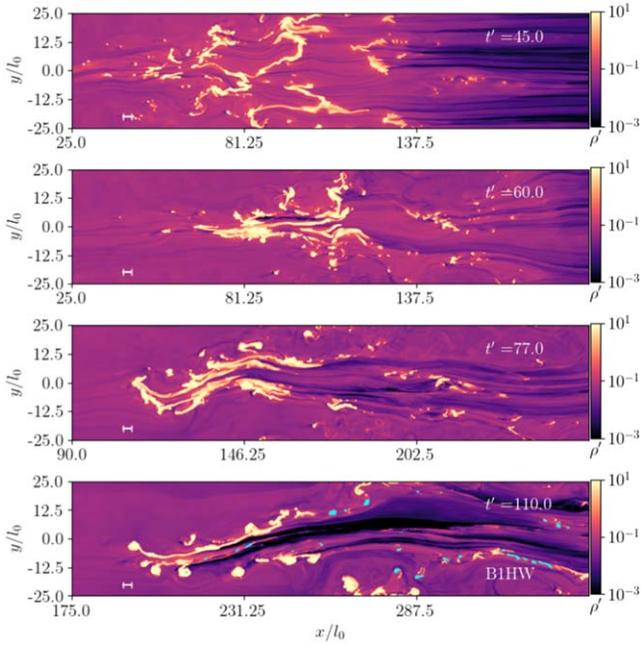

**Figure 3.** The gas density snapshots of B1HW, from top to bottom: $t' = 45$, 60, 77, and 110. The cyan masked regions in the last row are the small clouds we selected (see Section 4.2). The total cold gas fraction in the small clouds (masked by cyan in the last row) is ~5%. The white line segment in the lower-left corner shows five times of the initial Field length in the x-direction.

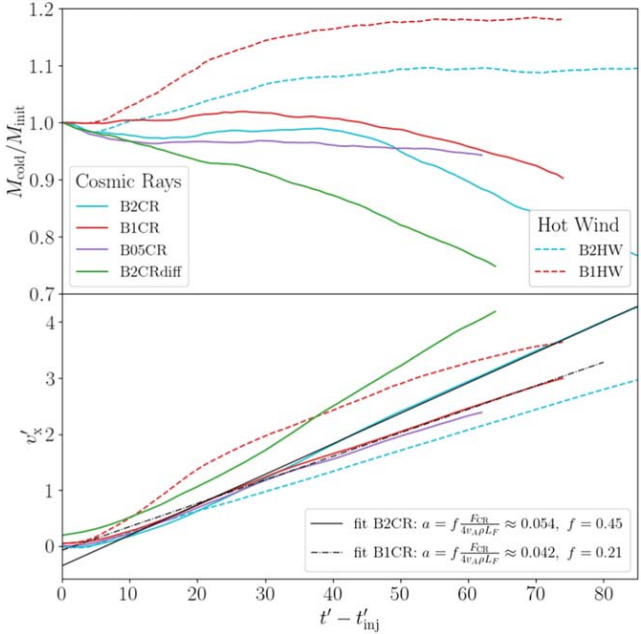

**Figure 4.** Upper panel: the mass evolution of cold gas for CR-driven runs (solid lines) and hot-wind-driven runs (dashed lines), with the mass scaled to the initial mass. The color notes the initial magnetic field or CR transport, cyan is B2CR and B2HW ($\beta_{init} = 0.5$), red is B1CR and B1HW ($\beta_{init} = 2$), purple is B05CR ($\beta_{init} = 8$), and green is B2CR diff ($\beta_{init} = 0.5$). Lower panel: mass-weighted average cold gas velocity in the x-direction. The black solid and dashed–dotted line are the linear fits according to Equation (10). We scale the start time to roughly when CR flux first interacts with the cloud at $t'_\parallel = 39$.

mass increases before $t' - t'_\parallel \sim 30$, and becomes more constant over time.

The cold gas growth is likely due to the efficient cooling of hot gas. Recent studies indicate that in multiphase gas, a thin, turbulent layer forms between the cold and hot phases where they mix (Fielding et al. 2020; Gronke & Oh 2020; Tan et al. 2021). The mixing layer is likely to be at the temperature that maximizes radiative cooling. The inhomogeneous and rapid cooling in turn redistributes pressure, potentially causing small-scale pressure gradients that further enhance turbulent mixing. As a result, the cold gas can grow due to the mixing-induced condensation of hot gas. Gronke & Oh (2018) proposed a framework for quantifying the cold gas mass evolution when entrained in a hot wind. It estimates the timescales for the two competing processes that dominate cold gas mass evolution: the compression and strong shock by the hot wind that destroys cold gas, and the radiative cooling of hotter gas that supplies the cold gas. To estimate the shock-induced cold gas destruction, we use the mass-weighted average density of the cold gas for $\rho'_c \approx 6.01$ to estimate the cloud-crushing time $t'_{cc} \approx 3.56$. Magnetic fields can modify the cloud-crushing process (McCourt et al. 2015) and potentially slow down cloud crushing. However, the magnetic field is relatively moderate in our simulation, so $t'_{cc}$ should provide order-of-magnitude estimation. The cold gas growth via condensation from hotter gas is characterized by the mixing layer cooling time $t'_{cool,mix}$. In their picture, the interface between cold and hot gas is a layer of intermediate-temperature gas with ongoing mixing. Their model estimates the mixing layer temperature $T'_{mix} \sim \sqrt{T'_c T'_h}$. For CGM-like values of cold gas with $T'_c \sim 10^4$ K and hot gas with $T'_h \sim 10^6$ K, $T'_{mix} \sim 10^5$ K, which is roughly the temperature that maximizes cooling. Following Gronke & Oh (2018), in B1HW and B1CR, we estimate the cooling time of the mixing layer as

$$t'_{cool,mix} = \frac{\rho'_c}{\rho'_h} \frac{\Lambda(T_{cold})}{\Lambda(T_{mix})} t'_{cool,cold} \approx 0.30 \quad (15)$$

where $T'_c \sim 0.20$ and $\rho'_c \sim 7.93$ are the mass-weighted average temperature and density of the cold gas, respectively, $T'_{mix} \sim 1.86$, and $t'_{cool,cold} \sim 0.01$ is the cooling time in the cold gas. In B1HW, $t'_{cool,mix} \ll t'_{cc}$, suggesting that the efficient cooling from the mixing layer is likely to compensate the cold gas destruction by dynamical processes and seed cold gas growth. Consistent with the estimation, the cold gas mass increases in B1HW.

### 3.3. Acceleration by Cosmic Rays: B1CR

Figure 5 shows the gas density snapshots of B1CR. Similar to B1HW, the initial clouds merge into clumps with sizes of a few initial Field lengths, but are then stretched in the x-direction. At late time, various fragmented small clouds are formed on the right side of the cold gas (for example, the clouds marked by cyan in the last row of Figure 5).

The bigger clouds are primarily pushed by the CR "bottleneck effect," which refers to the CR pressure gradient formed when CRs stream into high-density regions from low-density regions. The bottleneck arises when CRs stream at the local Alfvén velocity down the CR pressure gradient. When they enter dense, cold gas, the Alfvén velocity drops and CRs are slowed down. For static flow and magnetic field, $P_{CR} v_A^{4/3}$ is approximately constant at steady state (Breitschwerdt et al. 1991; Jiang & Oh 2018; Hin Navin Tsung et al. 2022). So more CRs pile up near the interface, increasing CR pressure, and the resulting CR pressure gradient can accelerate the cold, dense gas.

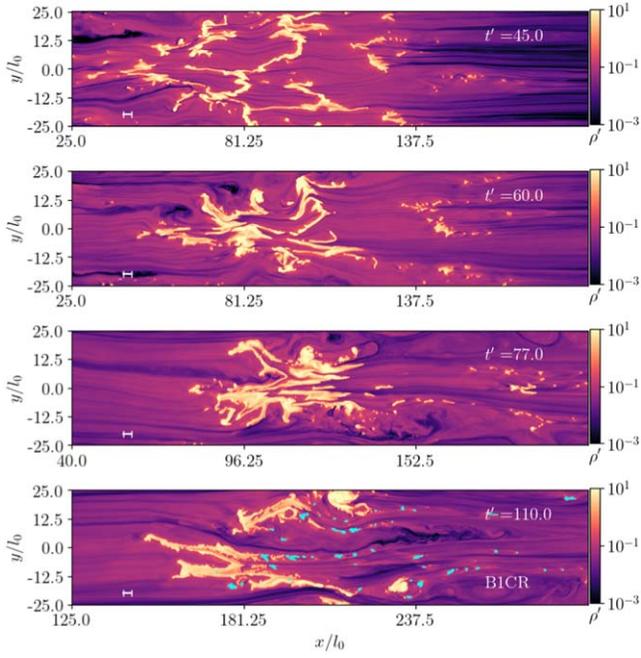

**Figure 5.** The gas density snapshots of B1CR, from top to bottom: $t' = 45, 60, 77, 110$. The cyan masked regions in the last row are the small clouds we selected (see Section 4.2). The total cold gas fraction in the small clouds (masked by cyan in the last row) is ~5%.

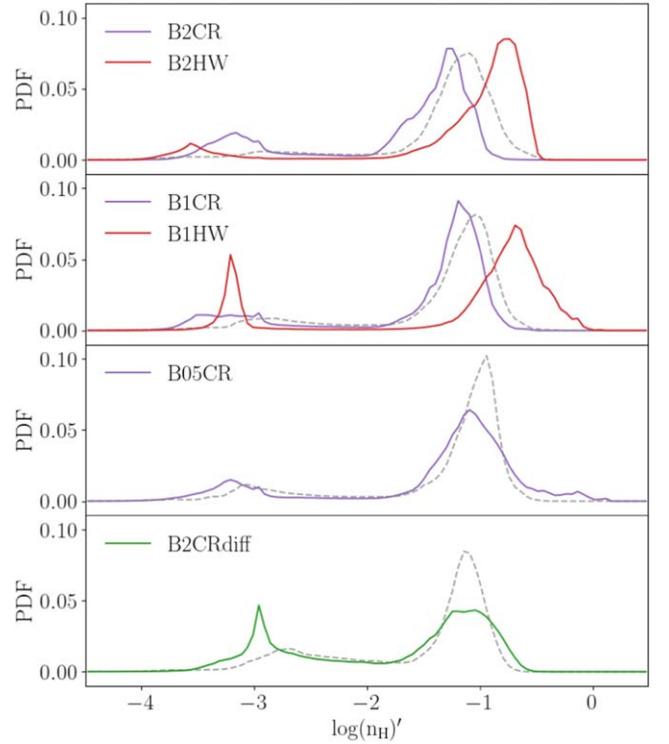

**Figure 6.** The mass-weighted density probability distribution of different sets of simulations sampled at specific time: B2CR and B2HW with initial $\beta = 0.5$ (the first row, $t' = 115$), B1CR and B1HW with initial $\beta = 2.0$ (the second row, $t' = 120$), B05CR with initial $\beta = 8.0$ (the third row, $t' = 90$), and B2CRdiff $\beta = 0.5$ (the fourth row, $t' = 90$). In each panel, the gray dashed line shows the distribution before injection of CR flux or hot wind; they are $t' = 30, 30, 30,$ and 26 from the first row to the fourth row.

The cold gas mass and mass-weighted average velocity evolution are shown in Figure 4 as the red solid lines. The acceleration of cold gas is nearly constant. We fit a straight line according to the constant acceleration estimated by Equation (10). Our results are consistent with $f \approx 0.21$, in agreement with the similar $f$ estimated in Brüggen & Scannapieco (2020), where a single, spherical cloud is accelerated by CR streaming.

The cold clouds in B1CR (Figure 5) are generally more diffuse and stretched than B1HW (Figure 3). The mass-weighted gas density distribution of B1CR and B1HW is shown in Figure 6, the two peaks corresponds to the $\sim 10^6$ K diffuse gas and $\sim 10^4$ K dense clouds. Compared to the initial distribution (gray dashed lines), acceleration by a hot wind enhances the density contrast, while acceleration by CR streaming preserves or slightly decreases the density contrast. If we estimate density contrast as the ratio between the two peaks, B2CR and B1CR yield $\rho_c/\rho_h \sim 80, 115$, while B2HW and B1HW has $\rho_c/\rho_h \sim 660, 340$.

The difference in density contrast is related to the driving mechanism. For entrainment in a hot wind, the cloud is driven by the thermal pressure shock formed at the interface between cold and hot gas. For CR streaming, the clouds are accelerated by the CR "bottleneck." Unlike the thermal shocks that preferentially compress the clouds, the CRs can penetrate into the cloud, providing nonthermal pressure support. When CR pressure in the cloud increases, the cloud needs a lower thermal pressure to reach pressure equilibrium with the surrounding hot gas. But the cloud temperature is constrained by the cooling function and already low, so it cannot drastically drop. Instead, the cloud stretches, resulting in lower cloud density and the elongated morphology. A similar reduction of cold gas density by CRs is observed in recent works (Wiener et al. 2019; Butsky et al. 2020).

In contrast to B1HW (red dashed line), the cold gas mass decreases in B1CR despite using the same initial condition and cooling function. In order to understand the cold gas drop, we consider the heating and cooling structure. Inside the dense, cold cloud ($T \lesssim 7 \times 10^4$ K) where the temperature is close to the floor temperature, cooling and heating are moderate. Surrounding the dense cloud, a thin layer of intermediate-temperature gas ($7 \times 10^4$ K $\lesssim T \lesssim 2 \times 10^5$ K) has the strongest cooling per volume. Around this thin cooling layer, the warm gas is dominated by a supplemental heating term $n_H\Gamma$. We note that it is possible that the heating constant $\Gamma$ is overestimated in our simulation. Before injecting CR flux or hot wind, the multiphase gas is in quasi-steady state, and the optical thin cooling term $n_H^2\Lambda$ roughly balances the supplemental heating. But supplemental heating in diffuse background gas is amplified after interacting with injected CRs or hot wind. We lower the heating constant by an order of magnitude in more diffuse gas with $\rho' \leqslant 0.1$ to minimize the effect from the nonequilibrium background after injecting CR flux or hot wind.

The thin layer of intermediate-temperature gas ($7 \times 10^4$ K $\gtrsim T_{gas} \gtrsim 2 \times 10^5$ K) around the cold gas dominates the volumetric emission, and the mixing and cooling in this layer strongly impact the cold gas mass evolution. Similar cloud emission structure is observed in Gronke & Oh (2020). Figure 7 shows the average density and total volume (area in two dimensions) of intermediate-temperature gas in B1HW and B1CR. Intermediate-temperature gas in B1CR has a larger total volume, but a lower average density than B1HW due to CR pressure support. The optically thin cooling is proportional to $n_H^2$, so B1CR has lower net cooling than B1HW despite the higher total volume.

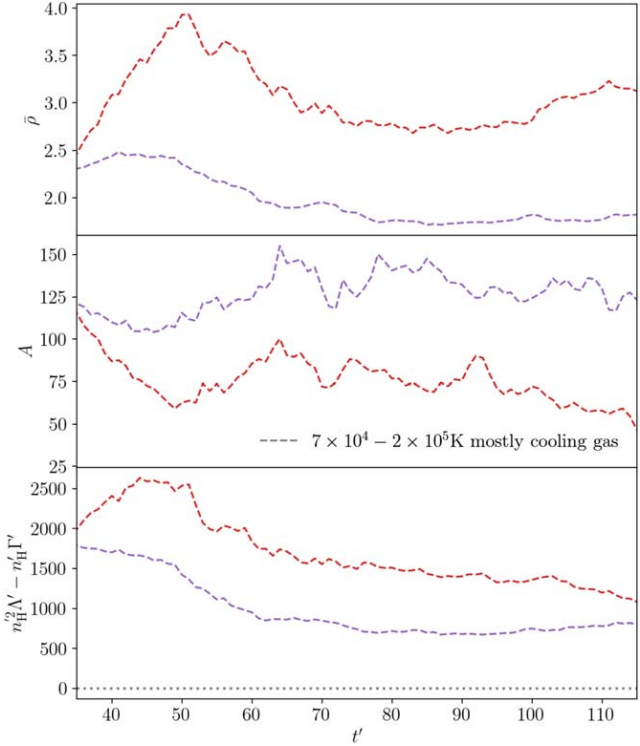

**Figure 7.** The average density (the first row), total volume (the second row), and total emission (the third row) for the intermediate-temperature gas (dashed lines) in B1CR (purple) and B1HW (red).

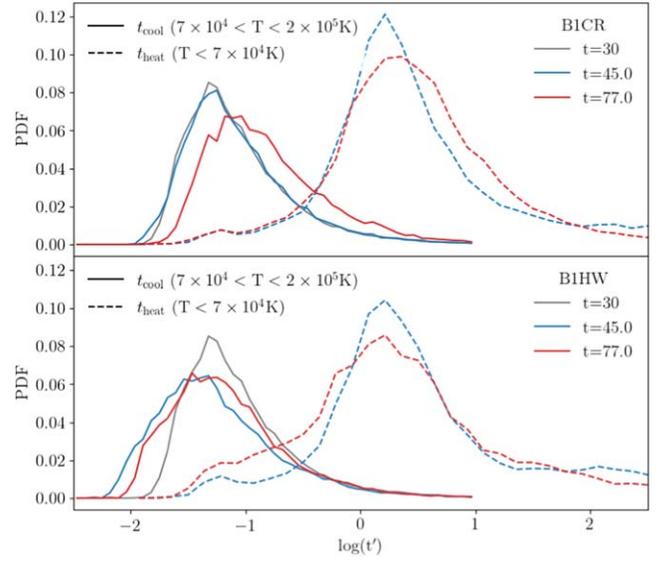

**Figure 8.** Probability distribution of cooling time for intermediate-temperature gas with $7 \times 10^4 \text{ K} \lesssim T_{\text{gas}} \lesssim 2 \times 10^5$ K (the solid lines) and heating time for cold gas $T_{\text{gas}} \lesssim 7 \times 10^4$ K (the dashed lines) for B1CR (the first row) and B1HW (the second row). The gray lines show the initial cooling time. During the acceleration, CR interaction with gas increases the cooling time while hot wind shortens the cooling time. The heating time of cold gas is usually long compared to the cooling time of intermediate-temperature gas.

The stronger cooling in B1HW also impacts the cooling timescale in the intermediate-temperature gas. Figure 8 shows the mass-weighted distribution of the intermediate-gas cooling timescale. Compared to the initial distribution (the gray lines), hot wind acceleration enhances the cooling rate and decreases the cooling time, in contrast to CR acceleration. Note, however, both the mass growth in B1HW ($\lesssim 20\%$) and the drop ($\lesssim 10\%$) in B1CR are relatively moderate given the fact that the duration is about $\approx 21 t'_{cc}$.

Meanwhile, the cold gas ($T \lesssim 7 \times 10^4$ K) is directly heated by both the supplemental heating of our net cooling function and the calculated CR heating. The corresponding heating time (the dashed line in Figure 8) is, however, long compared to the cooling time of intermediate gas, suggesting that the cold gas is not effectively heated. So the change of cold gas mass is more affected by cooling and dynamical processes.

Another factor that could influence cold gas evolution is the generation of small clouds at later times (marked by cyan in the last row of Figure 5). These smaller clouds are more likely to be destroyed in acceleration (Gronke & Oh 2018; Gronke et al. 2021), which should lead to cold gas loss. Using the method discussed in Section 4.2, we estimate the total cold gas mass in small clouds, but find that it is a small fraction for these fiducial runs. The generation of fragmented small clouds is observed in B1HW too with a similar low cold gas mass fraction. Thus, we conclude that the decreasing cold gas mass seen in B1CR is mostly the result of the suppressed cooling in the mixing layer.

### 3.4. Strong Magnetic Field: B2CR and B2HW

#### 3.4.1. Dynamics

We increase the initial magnetic field by a factor of two in B2HW and B2CR, giving an initial $\beta = 0.5$. The first column in the third row of Figure 2 shows the gas density snapshot before injecting CR or hot wind.

The right column of Figure 9 shows a series of the gas density snapshots from B2HW. The initially distributed cold clouds are pushed together by the thermal wind, and mostly merge into a single cloud, experiencing a collective acceleration. The cloud morphology differs from B1HW. The strong magnetic field constrains the fluid to be more laminar, and prevents the formation of spherical clouds.

B2HW also experiences total cold gas growth (see Figure 4). We use Equation (9) to estimate the cloud-crushing time $t'_{cc} \approx 4.90$, and the mixing layer cooling time is estimated to be $t'_{\text{cool,mix}} \approx 0.46$. We find $T'_c \sim 0.20$ and $\rho'_c \sim 6.52$ for the mass-weighted average temperature and density of the cold gas, $T'_{\text{mix}} \sim 1.66$. We find $t'_{\text{cool,mix}} \ll t'_{cc}$, consistent with the cold gas growth we observe.

The left column in Figure 9 shows the gas density snapshots of B2CR. The dynamics includes two major stages: when $t' \lesssim 75.0$, the initially distributed clouds are pushed by CRs and merge. After $t' \gtrsim 75.0$, the merged cloud is accelerated by CR streaming along the initial magnetic field direction.

Figure 10 shows the gas density, temperature, and pressure profiles across the merged cloud at $t' = 115.0$ for B2CR and B2HW. In B2CR, a CR pressure gradient is formed across the merged cloud (the bottleneck $225 \lesssim x' \lesssim 250$), and the gas pressure in the cloud is small. In B2HW, at $x' \sim 192$, thermal pressure forms a shock at the cloud interface and drops as one moves to the right within the cloud, while the CR pressure is unimportant. Notice that the gas pressure in B2HW and CR pressure in B2CR in hot gas on the left side are similar ($P' \sim 5.0$) due to the equivalent momentum flux supplied at the boundary in the two simulations. In both runs, magnetic pressure plays a nonnegligible role in the multiphase gas.

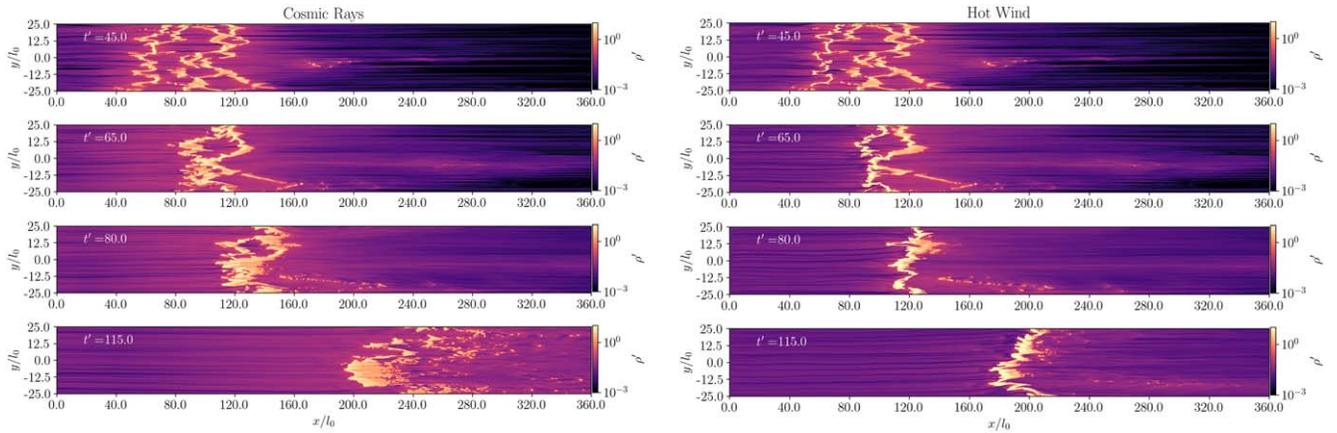

**Figure 9.** Gas density snapshots of CR-driven runs (the left column) and hot-wind-driven runs (the right column). From top to bottom, the snapshots are taken at $t' = 45.0, 65.0, 80.0,$ and $115.0$.

Dynamically, the cloud merging is an important process that modifies the CR pressure distribution in our simulations. Figure 11 shows a typical cloud merging process. At $t' = 50$, three CR "bottlenecks" forms when encountering three clouds (the gray filled region in the first row) at $x' \sim 72, 96,$ and $136$. The CR pressure jump usually occurs at a local minimum of $v'_x + v'_A$. As the clouds merge and CRs penetrate into the clouds, the bottlenecks are "merged" too. At $t' = 63$, CRs roughly decouple with gas in the region $x' \lesssim 110$, resulting in nearly flat CR pressure despite the local density variation. The stair-like CR pressure profile present with multiple moving clouds and final merged cloud profile is consistent with the theoretical study of CR bottlenecks in Hin Navin Tsung et al. (2022).

After cloud merging, the cold gas experiences a collective bulk acceleration. During acceleration, we observe that multiple small-scale clouds are formed on the right side of the merged cloud (the fourth row in Figure 9). These clouds are usually smaller than the initial Field length, and tend to quickly mix with the hot background gas. Similar small clouds are also observed in B1CR at later time. But the effect seems to be enhanced in B2CR. We discuss the formation and destruction of these small clouds in Section 4.2.

The first row of Figure 6 shows the density distribution in B2CR and B2HW. Similar to B1CR and B1HW, the density contrast is reduced in B2CR due to CR pressure support, and enhanced in B2HW due to the thermal pressure compression.

The total cold gas mass and average velocity of B2CR is plotted in Figure 4. The acceleration is also linear and is well fit by Equation (10). The average Alfvén velocity is higher than B1CR due to the stronger magnetic field, but $f \approx 0.45$ is also higher than B1CR, resulting in similar acceleration.

The higher $f$ seems to be a result of a larger effective area. The merged clouds in B2CR are not significantly compressed in the y-direction and maintain the initial covering fraction. The strong magnetic field alignment is not significantly altered by thermal instability (see also Jennings & Li 2021), and CR streaming is relatively uniform in the y-direction in B2CR and B2HW, so that the clouds maintains the "shell" morphology, and thus larger effective area. This differs from B1CR, where the magnetic field is distorted by thermal instability. CRs preferentially stream through the channels between the cold clouds, resulting in a CR pressure gradient in the y-direction that compresses the merged cloud and reduces the effective area.

### 3.4.2. Cold Gas Evolution

B2CR experiences cold gas loss during the acceleration while B2HW has cold gas growth, consistent with what we observed in B1CR and B1HW. We find that in B2CR and B2HW, however, the intermediate-temperature gas has a similar cooling timescale in the two runs. The cooling timescale distribution has a similar shape to B1CR (see Figure 8), with the peak shifted to $\log(t_{\text{cool}})' \sim -1.1$. We subtract all external thermal sources including supplemental and CR heating when calculating the net cooling timescales. The effect of CR heating $\sim v_A \cdot \nabla P_{CR}$ is generally unimportant in the intermediate-temperature gas. We also find that heating of cold gas seems to be unlikely to drive the observed mass change in B2HW and B2CR.

The similarity of intermediate-temperature gas arises from the pressure support. Figure 10 shows that the magnetic pressure is significant in both runs. On the right side of the cloud, opposite the incoming CRs or hot wind, gas is supported by magnetic pressure. The purple band in the temperature plot labels the intermediate-temperature gas, which occupies a larger volume on the right side of the cloud too. So the intermediate-temperature gas density is primarily set by the magnetic pressure, which is similar in B2HW and B2CR.

Figure 12 shows the total emission, emission from the intermediate-temperature gas, and the rate of change of cold gas in B2CR and B2HW. Because the intermediate-temperature gas has similar properties between the two simulations, the emission in the two runs is also similar (dashed lines in the lower panel). The thin solid lines show the total emission from all gas. Interestingly, in B2HW, the emission from intermediate-temperature gas (dashed red line) traces the total emission (thin solid red line) well, suggesting that the net heating and cooling from the gas with other temperatures in the simulation roughly balance each other. In B2CR, the intermediate-temperature gas (dashed purple line) emission is similar to hot wind, but the negative overall emission (thin solid purple line) suggests that the gas with other temperatures experiences net heating. We found that the net heating in B2CR is due to a larger volume of warm gas ($\sim 10^{5.3-6}$ K) compared to B2HW. The warm gas usually surrounds cold clouds and forms from mixing, and it is dominated by supplemental heating. The CR-supported cold clouds have larger surface area, where mixing happens, leading to a higher volume of warm gas, and thus, total heating.

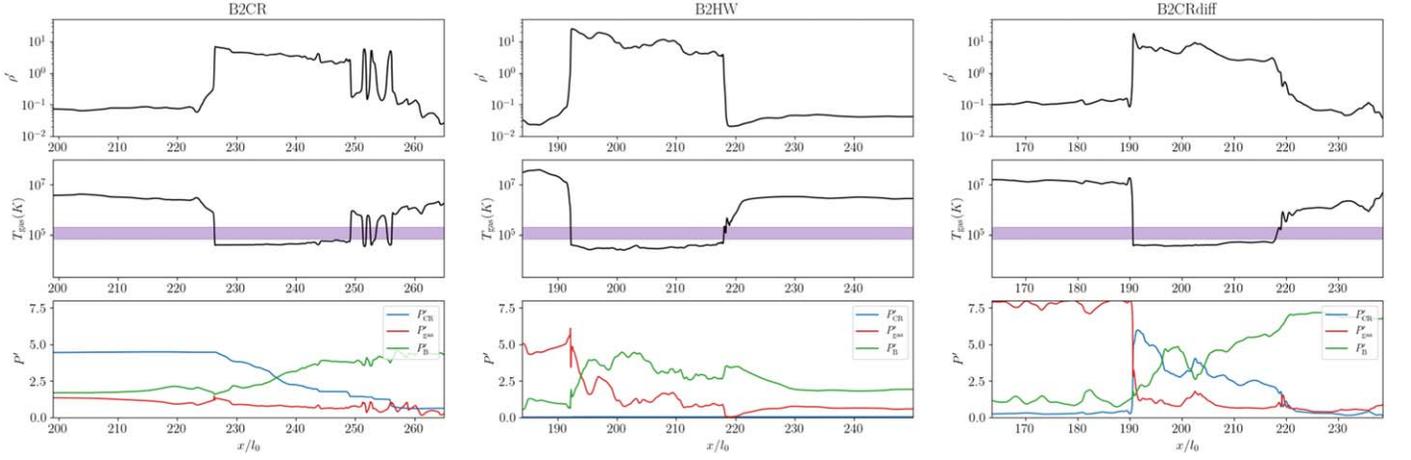

**Figure 10.** Representative merged cloud structure at late time in B2CR, B2HW, and B2CRdiff. The density (the first row), temperature (the second row), and pressure (the third row) are profiles of a line cut through the gas from $x' = 119.0$ to $x' = 265.0$ at $y' = -16.5$, for B2CR (the first column); $x' = 119.0$ to $x' = 265.0$ at $y' = -15.7$ B2HW (the second column); and at $t' = 115$ and B2CRdiff from $x' = 163.5$ to $x' = 238.5$ at $t' = 90$ (the third column). In the third row, the blue solid line is the CR pressure, the red solid line is gas pressure, and the green solid line is magnetic pressure. The purple band in the second row labels the intermediate-temperature gas ($7 \times 10^4$ K $\lesssim T_{\rm gas} \lesssim 2 \times 10^5$ K).

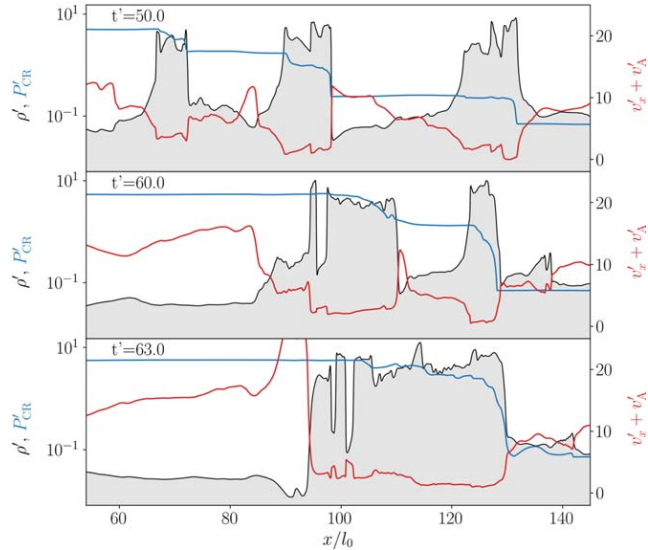

**Figure 11.** The time series of gas density (the black solid line and shaded region), CR pressure (the blue solid line), and the sum of flow velocity and Alfvén velocity (the red solid line) at $y' = -11.4$ at $t' = 50$ (the first row), $t' = 50$ (the second row), and $t' = 63$ (the third row) from B2CR.

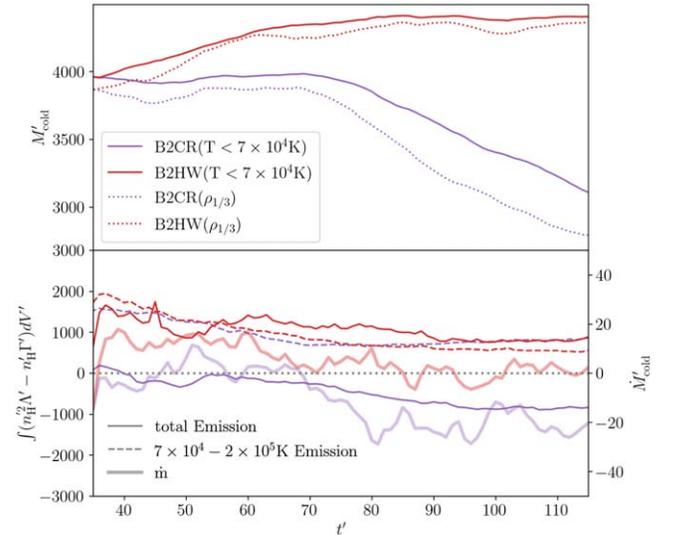

**Figure 12.** The cold gas mass and emission as a function of time. Line color denotes the driving mechanism: the purple lines are for CR-driven, and the red lines are for hot-wind-driven runs. Upper panel: the cold gas mass evolution. The solid lines are defined as gas with temperature $T_{\rm gas} < 7 \times 10^4$ K, while the dotted lines are defined as gas with density lower than $1/3 \bar{\rho}$, where $\bar{\rho}$ is the mean density. Lower panel: thin solid lines and dashed solid lines with labels on the left are the net emission from gas. The thin solid lines are total emission from the simulation domain, while the dashed lines are the emission from intermediate-temperature gas. The thick solid lines, with labels on the right, are the derivatives of the cold gas mass (the solids lines in the top panel).

If the cold gas evolution is primarily determined by the cooling and heating, the total emission should be roughly proportional to the cold gas mass rate of change (Fabian 1994). The lower panel of Figure 12 shows that the total emission (the thin solid lines) strongly correlates with the cold gas mass rate of change (the thick solid lines), but deviates from the proportionality in the acceleration stage ($t \gtrsim 70$ for B2CR), suggesting that there are other factors affecting the cold gas evolution.

Figure 9 shows that at later time, B2CR produces multiple small clouds on the right side of the merged cloud. The small clouds mix more easily with the hot background gas or are destroyed by deformation. We estimate that in B2CR, $\sim$20% of the cold gas is in the form of small clouds, while the fraction is usually $\lesssim$5% in B2HW. The morphological difference seems to explain the mass loss in B2CR. We discuss the effect of small clouds in further detail in Section 4.2.

### 3.5. Weak Magnetic Field: B05CR

We decrease the initial magnetic field by a factor of two in B05CR, given initial $\beta = 8.0$. The second column in the third row of Figure 2 shows the gas density snapshot before injecting CRs. The overall dynamics of B05CR is similar to B1CR. Thermal instability creates more fragmented small clouds scattered at $x' \gtrsim 112.5$. We observe a similar stretched cloud morphology during the acceleration. The average cold gas mass and velocity is similar to B1CR, too. Figure 13 shows that the magnetic field, however, is more turbulent in this simulation. The initial uniform magnetic field strength is $B'_{\rm init} = 0.5$. The upper panel shows the snapshot before we inject the CR flux, and the magnetic field strength in the cold gas has already been enhanced by thermal instability. In the lower panel, after CRs enter the domain, their interaction with multiphase gas further modifies the magnetic field, and the magnetic pressure in cold gas can be an order of magnitude larger than the initial value.

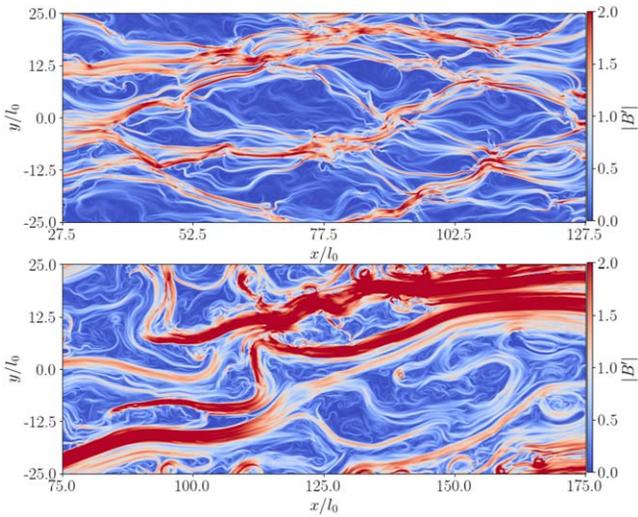

**Figure 13.** The magnetic pressure in B05CR before injecting CR flux ($t' = 30$, the upper panel) and after the clouds are accelerated by CRs ($t' = 75$, the lower panel). Notice that we show the magnetic pressure where most of the cold gas is located, so these two panels has different $x$-coordinate ranges.

CRs stream along the entangled magnetic field, resulting in a CR pressure gradient in both the $x$- and $y$-directions. In addition, $B'_y$ is amplified, making $v_{\rm A,y}$ comparable to $v_{\rm A,x}$ in some of the hot gas. In these regions, the CR pressure gradient in the $y$-direction is also comparable to the $x$-direction, leading to additional CR heating in the $y$-direction ($\sim v'_{\rm A,y} \partial_y P'_{\rm CR}$). The cooling in the background gas is generally inefficient, and the CR heating can not be quickly radiated away. The heating time of this gas $t'_{\rm heating} \sim 0.03$ can be one to two orders of magnitude smaller than those in strong magnetic field runs. As a result, gas pressure in the original $\sim 10^6$ K gas is increased, cold gas is accelerated by both gas and CR pressure. For the $x$-direction acceleration, the average fractional contribution from the enhanced gas pressure is $\sim 5.2$ times the CR pressure. In the turbulent magnetic field, CR streaming drives turbulence and heats the gas. CRs alone are relatively inefficient to accelerate gas in a preferred direction.

Although cold gas is not primarily accelerated by CRs, the intermediate-temperature gas is still supported mainly by CR pressure. So the intermediate-temperature gas properties, cooling timescale, and cold gas rate of change are similar to those of B1CR. In the later period of acceleration, CRs penetrate into the cold gas, and the cloud is again supported by CR pressure so that the density contrast of the multiphase wind is similar to B1CR.

### 3.6. The Effect of Conductivity

We now explore the effect of lowering the conductivity, which allows us to have a lower value of the Field length at fixed density or keep the Field length fixed while lowering the density. In B2CR_LC, we lower the conductivity ($\kappa'_\parallel$ and $\kappa'_\perp$) by a factor of 100. This allows the physical value of the Field length, and therefore cloud size, to remain constant while lowering the density by a factor of 10. This also corresponds to a factor of 10 reduction in the heating constant $\Gamma$ as described in Section 2.3. Hence, the motivation for B2CR_LC is to explore the relative impact of the heating constant, while keeping cloud sizes to be a few hundred parsecs.

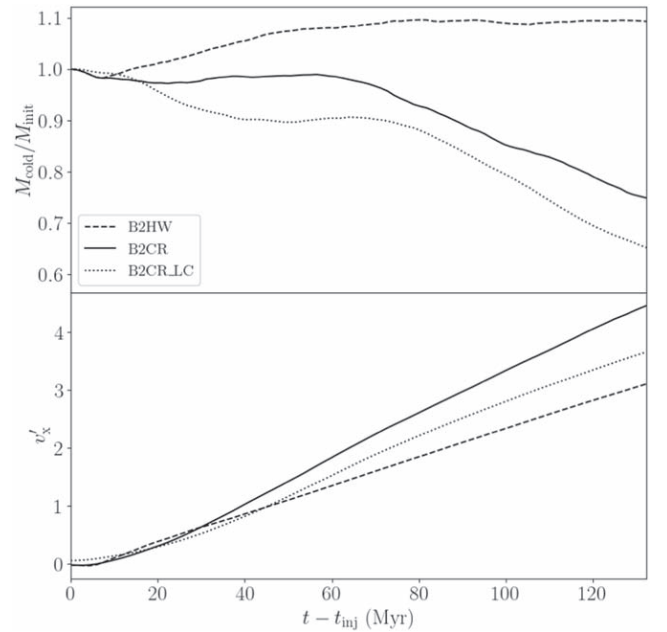

**Figure 14.** Similar to Figure 4. Upper panel: the mass evolution of cold gas B2CR (solid lines), B2HW (dashed lines), and B2CR_LC (dotted lines). Lower panel: mass-weighted average cold gas velocity in the $x$-direction. We scale the start time to roughly when the CR flux first interacts with the clouds.

We only run the thermal instability to $t' = t'_\parallel = 10$ in B2CR_LC, as formed multiphase gas reaches a quasi-steady state faster than B2CR due to the lower conductivity. We inject $F'_{\rm CR} = 60$ into the domain from the left boundary as in B2CR, so that the CR pressure is similar relative to the gas pressure and magnetic pressure.

The overall dynamical process is similar to B2CR despite some morphological differences. The cold clouds initially formed from thermal instability are less clumped than those in B2CR, but the cloud sizes are still comparable to the Field length. After interacting with CRs, the clouds are slightly less merged compared to the bulk cloud in B2CR. This is likely due to the smaller cloud column density. When accelerated by CRs, the clouds are also stretched in the $x$-direction, and several small clouds are formed in the acceleration. Similar to B2CR, we found a noticeable fraction of cold gas ($\sim 25\%$ at late time) is in the form of small clouds.

Figure 14 shows the cold gas mass evolution and acceleration of B2CR_LC. The cold gas evolution is similar to B2CR. We find that the cooling and heating structure is not sensitive to the lower conductivity, density unit, and heating constant. The intermediate-temperature gas is still a thin layer between the cold and hot gas that emits the most per volume. The warm gas ($\sim 10^{5.3-6}$K) is also still dominated by supplemental heating despite the order of magnitude reduction in the heating constant. In addition, we do not observe significant growth of the thin intermediate-temperature gas layer, and the average density of the gas is similar to B2CR. Interestingly, at late times, the magnetic pressure has a smaller fractional contribution to support the intermediate-temperature gas compared to B2CR. The lower magnetic pressure in intermediate-temperature gas might be related to both the different magnetic field distortion in the thermal instability stage and the lower level of merging that occurs in the acceleration stage.

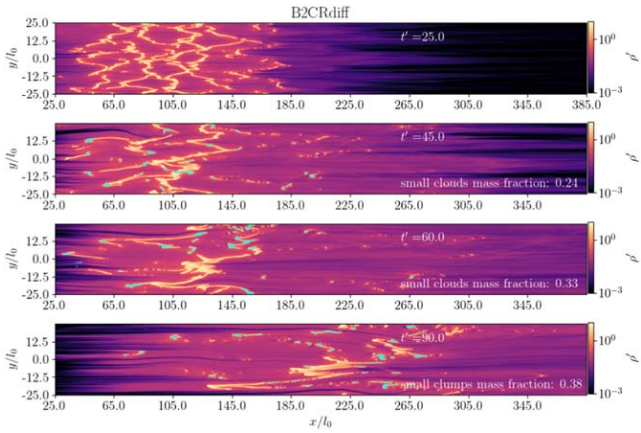

**Figure 15.** Gas density snapshots from B2CRdiff, which, from top to bottom, are taken at $t' = 25.0$, 45.0, 60.0, and 90.0. In the last row, the blue masked regions are the small cold clouds with temperature $T < 7 \times 10^4$ K. The small clouds fraction is about ~25%.

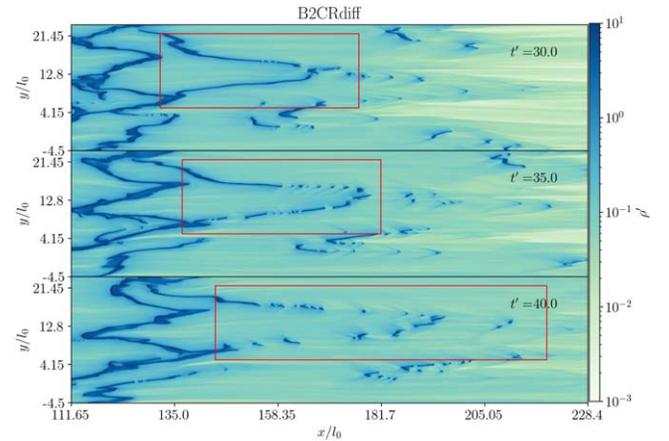

**Figure 16.** Zoomed-in density snapshots of B2CRdiff at $t' = 30$, 35, 40, and 45 (from the top to bottom panels). The snapshots show that the filament structure (highlighted by the red rectangle) at $t' = 30$ breaks into several small clouds distributed between $158 \lesssim x' \lesssim 228$ from $t' = 30$–45.

### 3.7. CR Diffusion

In B2CRdiff, we assume that CR transport is diffusive and we turn off CR streaming ($v_s = 0$). The diffusivity $\kappa_{\rm diff} = 1.73 \times 10^{26}$ cm$^2$ s$^{-1}$ is constant, giving the diffusion timescale $t'_{\rm diff} \sim L_F^2/\kappa_{\rm diff} \approx 10.27$, about two orders of magnitudes longer than the initial cooling time. The long diffusion timescale means that CRs travel slowly through gas, in contrast to the relatively fast streaming in B2CR. This choice of $\kappa_{\rm diff}$ is smaller than what is usually considered likely to occur in realistic systems. Hence, a shorter diffusion time may be expected in more realistic systems.

Figure 15 shows the density snapshots of B2CRdiff. The clouds are stretched in the $x$-direction and accelerated without significant merging. This produces a different cloud morphology from streaming, with clouds connected by filaments along the magnetic field direction, which is consistent with the findings in Sharma et al. (2010b).

Figure 6 shows that the accelerated clouds roughly maintain their initial density contrast in B2CRdiff. The third column in Figure 10 shows the density, temperature, and pressure of a typical stretched cloud in B2CRdiff. The diffusion time in our simulation is substantially longer than other dynamical timescales, and CRs are trapped in the dense cloud, giving a high CR pressure in cold gas. When CR flux enters the domain, it compresses the warm and background gas, increasing the gas pressure. The high gas pressure forms a shock at the cloud interface and accelerates the cold gas. B2CRdiff has somewhat higher acceleration than B2CR, which we attribute to this increasing gas pressure and the dropping total cold gas mass.

The overall higher density leads to a shorter cooling time in the intermediate-temperature gas. This short cooling time, however, does not increase the cold gas mass. In fact, B2CRdiff has the largest fractional cold gas loss among all of the simulations. We attribute the loss to the production of small clouds and their efficient destruction in B2CRdiff. In Figure 15 we denote the small clouds with cyan masked regions. At $t' = 45$, 60, and 90, the fractions are ~24%, 33%, and 38%.

The mechanism for generating small clouds is also different from that in the CR streaming case (B2CR). A noticeable fraction of small clouds are formed at both sides of the shell, even at early time. We find that they primarily originate from the breakup of filaments. Figure 16 shows a typical small cloud generation process: a filament (noted by the red rectangle) at $t' = 30$ breaks into several small clouds. We further discuss the formation of small clouds in Section 4.2.

Diffusion and streaming operate intrinsically differently when creating and accelerating multiphase gas. Sharma et al. (2010b) suggested that the filamentary structure is a natural result of thermal instability with anisotropic conductivity. The fastest growing mode is parallel to the magnetic field, in which the conductivity is optimal, resulting in filaments along the magnetic field. This morphological difference of the resulting clouds leads to different cold gas evolution.

## 4. Discussion

### 4.1. CR Streaming in Nonuniform Magnetic Field

Figure 13 shows that the initial weak magnetic field can be amplified by thermal instability and CR streaming. In B05CR, the initial $\beta_{\rm init} = 8$. Before injecting CRs, the average $\beta$ of cold gas is amplified by thermal instability to $\bar{\beta}_{\rm cold} \approx 1.5$. At $t' = 90$, CR acceleration further enhances magnetic pressure in the cold gas to $\bar{\beta}_{\rm cold} \approx 0.75$, which is reduced by about an order of magnitude compared to the initial value. The interaction also entangles field lines, resulting in a highly nonuniform and dynamical magnetic field. Although CR interaction with a cold cloud in a relatively uniform magnetic field is investigated by recent works (Wiener et al. 2019; Brüggen & Scannapieco 2020), CR streaming in nonuniform magnetic fields is less studied. In this section, we discuss the CR interaction with multiphase gas in the nonuniform magnetic field created by thermal instability in our simulations.

When magnetic field lines are well-aligned with the direction of incoming CR flux ($x$-direction), CR pressure shows the classical "bottleneck" that drives cold clouds. When the field lines are entangled, CR streaming does not directly provide strong acceleration in the $x$-direction. But how are CRs distributed along the magnetic field line? The lower panel of Figure 17 shows the CR pressure along a distorted magnetic field line (the black solid line in the upper panel) in B05CR. Interestingly, the CR pressure and gas density show a typical "bottleneck" (similar to Figure 11). The red line shows the magnitude of the total CR streaming velocity in the lab frame along the magnetic field $|\mathbf{v}'_\parallel + \mathbf{v}'_{A,\parallel}| = (\mathbf{v}' + \mathbf{v}'_A) \cdot \mathbf{e}_\parallel$, where $\mathbf{e}_\parallel$ is the unit vector tangent to the field line. However, when we project the CR pressure gradient onto the $x$-direction, the acceleration is small for two reasons. First, the magnetic field is highly nonuniform. In some regions, the enhanced magnetic field in the colder gas compensates for the high density, making the $v_A$ comparable to nearby hotter gas. So the "bottleneck" is reduced and the CR pressure gradient is shallower. Second, the turbulent magnetic field is inefficient in transporting CRs preferentially along the $x$-direction. At early time $t' \lesssim 60$, CR pressure in the cold gas can be even lower than the gas pressure. The smaller CR pressure and shallower pressure gradient constrains CRs' ability to directly drive cold gas.

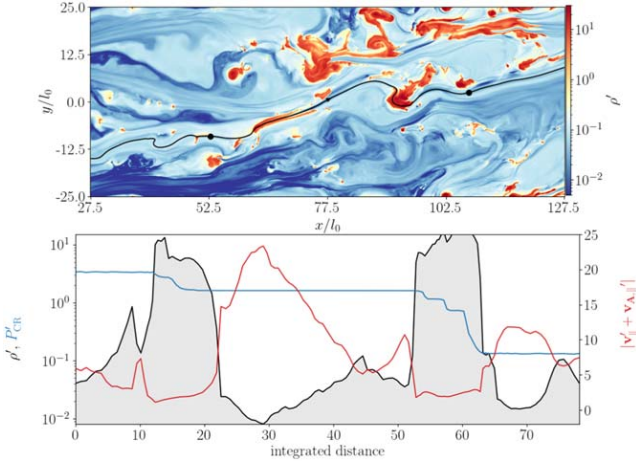

**Figure 17.** Upper panel: gas density from B05CR at $t' = 65$. The black solid line labels a typical magnetic field line, and the two black dots shows the start and end points of the segment we plot in the lower panel profiles. Lower panel: the gas density (black solid line and gray shades), CR pressure (the blue solid line), and total projected velocity.

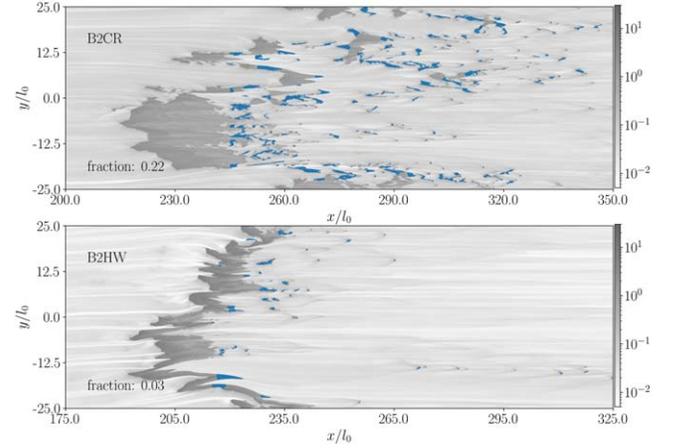

**Figure 18.** Small clouds (masked by blue) identified by Watershed in B2CR (upper panel) and B2HW (lower panel) at $t' = 115$. The fractions of cold gas mass in the small clouds are $\sim 21\%$ in B2CR and $\sim 5\%$ in B2HW.

Nevertheless, CRs are an interesting source of energy and momentum with a turbulent magnetic field. As discussed in Section 3.5, B05CR produces the most CR heating among the three CR streaming simulations. The turbulent magnetic field helps CR heating by amplifying $v_{A,y}$ and providing CR pressure gradient in both directions. The hotter gas plays an important role in accelerating the colder gas. From a momentum point of view, CRs streaming along a turbulent magnetic field increase the velocity dispersion in both $\sim 10^4$ K and $\sim 10^6$ K gas. In B2CR and B1CR, the velocity dispersion of $\sim 10^4$ K gas $\sigma_v$ peaks in the cloud merging stage and drops during the collective acceleration. However, $\sigma_v$ constantly grows in B05CR. At $t' = 90$, $\sigma_v \approx 80$ km s$^{-1}$ in B05CR, which is about a factor of two larger than B1CR and B2CR.

### 4.2. The Generation and Destruction of Small Clouds

In B2CR and B2CRdiff, the generation of small clouds is an important process for driving the cold gas loss seen in our simulations. In B2CR, part of the cloud detaches from the merged cloud structure and further breaks into smaller pieces. In B2CRdiff, the filaments break into multiple small clouds. In this section, we discuss their formation and potential impact on cold gas survival.

We find that the small clouds represent a nonnegligible fraction of cold gas mass in B2CR and B2CRdiff. We tracked the evolution of small clouds in the simulation and found almost all of them are mixed with the hot background and destroyed within $t'_{cc}$, contributing to the observed cold gas loss. In order to quantify the cold gas fraction in the form of small clouds, we use the image segmentation algorithm Watershed from scikit-image (Van der Walt et al. 2014). A similar technique is adopted in studies with morphology-based image segmentation (Lin et al. 2016; Krieger et al. 2021). For example, Figure 18 shows the density snapshots at $t' = 115$ from B2CR and B2HW. The regions masked by blue are the selected small clouds.

We use the subdomain of $245 < x' < 320$ for B2CR and $215 < x' < 295$ for B2HW to apply Watershed. First, we make a mask by selecting gas with $T_{gas} \leqslant 7 \times 10^4$ K. Then we obtain the Euclidean distance transform of the mask array (by using the distance_transform_edt function). Next, we find the local maximums of the distance array and use their coordinates as the seed for Watershed. The minimum separation of local maximum peaks is set to be 40 pixels, but we adjust it when calculating other density snapshots to obtain the optimal segmentation. Finally, we select out the segmented regions with total area less than $A'_{tot} \leqslant A_{cut}$. We tested $A_{cut} = 7.5$, 10, and 11.25 and found our results are not sensitive to this choice, adopting $A_{cut} = 10$ for the results shown here. Some regions within the big cloud are mis-selected due to oversegmentation. Thus, we tested different parameters in the above selection pipeline, or manually excluded the mis-selected regions. At $t' = 115$, the cold gas mass in small clouds is about $\sim 22\%$ in B2CR, and only $\sim 3\%$ in B2HW. We also checked the snapshot at $t' = 85,100$, and the fractions are about $\sim 17\%$ and $\sim 19\%$ in B2CR and $\sim 3\%$ in B2HW. An important consequence is that these small clouds are easier to destroy by mixing in later dynamical process, which is consistent with the evolution we observe in the simulations.

In contrast, the small clouds are a less important factor in B1HW, B1CR, and B05CR (see, e.g., Figures 3 and 5). The fraction of cold gas in small clouds at $t' = 110$ is ∼5% for B1CR and ∼5% for B1HW. We also checked the cold gas fraction in small clouds at $t' = 77$ and $t' = 90$, and found it to be ∼5% and ∼4% in B1CR, and ∼5% and ∼5% in B1HW, respectively. A similar low fraction is found in B05CR at multiple times, ∼7% for $t' = 65$ and $t' = 95$.

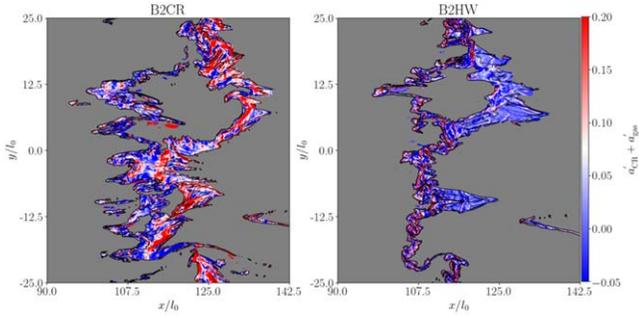

**Figure 19.** The sum of CR acceleration $a'_{CR} = \sigma'_{c,xx}(F'_{CR} - 4v'_x P'_{CR})$ and thermal pressure acceleration $a'_{gas} = -\partial P'_{CR}/\partial x'$ in the $x$-direction. Gas with temperature $T \geqslant 10^5$ K is masked by gray, and the black lines draw contours for $\rho' = 2.5, 10$. CRs or a hot wind comes in from the left side, so that B2CR (left) experiences stronger acceleration on the far side while B2HW (right) experiences stronger acceleration on the incoming side.

The small cloud generation is related to CR–gas interaction. In B2CR, small clouds are created when cold gas detaches from the merged clouds. We find that the differential acceleration by CR bottleneck is the primary reason. For example, Figure 11 shows that when the magnetic field is relatively uniform and the Alfvén velocity almost solely depends on density, the bottleneck is strongest when CRs exit the cold clouds. At the incoming side, CRs decouple from gas and stream freely at the reduced light speed. At the exiting side, the gas density gradient is the same as the streaming direction, CRs re-couple with gas and stream at $v_A$, creating a CR pressure gradient. Hence, the far side usually experiences stronger acceleration.

Figure 19 compares the total acceleration from CR and thermal pressure on cold gas. As explained, B2CR has the strongest acceleration on the far side. This gas constantly experiences enhanced CR acceleration until the CRs decouple from the gas. The irregular shape of the clouds also contributes to the nonuniform acceleration following detachment. After the fast-moving cloudlets leave the merged bulk, they fragment into smaller clouds. In contrast, B2HW has the strongest acceleration on the wind incoming side, the net effect is compression of the cloud, and the far side gas is free from excessive acceleration.

The effect also depends on the magnetic field. In B2CR, the differential acceleration location is relatively fixed in the comoving frame, which is essential for stripping clouds from the bulk. However, in weaker magnetic field simulations, the field lines are not well-aligned along the $x$-direction; CRs stream in both the $x$- and $y$-directions. Generally, the cold gas motion is more turbulent, so the location of the maximum CR pressure gradient changes in the comoving frame. The differential acceleration and subsequent cloud fragmentation are less significant in B1CR, B1HW, and B05CR.

When streaming is turned off and "bottleneck" disappears in B2CRdiff, the breaking filaments are the main source of small clouds. Due to their elongated morphology, filaments are fragile to even moderate nonuniform acceleration. Since these filaments are created by thermal instability, cold gas loss starts as soon as the diffusive CR flux interacts with the gas

(Figure 4). This differs from B2CR, where the differential acceleration happens after the cloud merging, so the drop starts at a later time. Nevertheless, as shown in Figure 15, B2CRdiff also has a significant fraction of cold gas in the form of small clouds despite the morphological difference, with ∼24%, ∼33%, and ∼38% at $t = 45$, 60, and 90.

Since the rate of production of small clouds impacts the rate of cold gas destruction, it is important to note that some of the smallest clouds in B2CR are only resolved by a few cells, and therefore may be destroyed too quickly in the simulation. To assess this, we estimate their final fate based on theoretical considerations. Gronke & Oh (2018) and Gronke et al. (2021) proposed that cold gas will survive in the acceleration if the cooling is fast compared to cloud crushing and following mixing. We follow their scheme to estimate the fraction of small clouds that are unlikely to survive in later acceleration (see also Section 3.4).

For each selected small cloud, Watershed also calculates the major axis length and eccentricity of the centroid ellipse with equivalent area. We retrieve the approximate diameter, average density, and temperature of each selected small cloud to estimate the cooling timescales and destruction timescale. In B2HW, we compare $t'_{mix,cool}$ with the cloud-crushing time $t'_{cc}$. In B2CR, we compare $t'_{mix,cool}$ with an analogy to a cloud-crushing time $t'_{cc}$, where we replace $v_{hot}'$ by the relative velocity of the cloud and background hot gas. For multiple time snapshots, we estimate that roughly ∼40% of the small clouds are inefficiently cooled and likely to be destroyed. This potentially leads to less cold gas loss than what is observed in B2CR if all small clouds are well resolved.

In summary, nonuniform acceleration shreds the merged cloud structure or breaks the filaments, redistributing cold gas into smaller clouds. The destruction of these small clouds contributes to the cold gas loss in B2CR ad B2CRdiff. Although some small clouds are not well resolved in our simulations, and the exact cold gas loss in this channel might depend on resolution, we estimate that a large fraction of the mass in these cold clouds will not survive.

### 4.3. Comparison to Previous Works

*Radiative Mixing Layer:* In Section 3.4.2, we showed that the thin layer between cold and hot gas dominates the volumetric cooling, and the intermediate-temperature gas and their emission in this layer strongly impacts the evolution of cold gas. Tan et al. (2021) and Fielding et al. (2020) studied the detailed structure of the interface between cold and hot gas. Assuming gas enters the mixing layer and can be quickly cooled, the inflowing velocity to the mixing layer is then a key quantity to deciding the cold gas growth rate. These works derive scaling relations for the mass flux into the turbulent radiative mixing layer, providing a useful quantification of the emission and cold gas growth rate in multiphase gas. Analogous to thermal combustion, Tan et al. (2021) proposed that the mixing layer can be characterized by the ratio of cooling timescale and mixing timescale. When cooling is fast compared to mixing, the interface gas will be fragmented and form a multiphase structure.

We observe a similar cooling-efficient, multiphase region at the cloud interface in our simulations. Consistent with their study, the local velocity dispersion maximum is cospatial with the strongest emission per volume, and the Damköhler number generally divides the single-phase and multiphase regions.

Nevertheless, when we estimate the velocity of mass flux into the turbulent mixing layer for the two sets of CR and thermal wind simulations (B1CR-B1HW and B2CR-B2HW), we do not obtain a mass flux difference that can explain the diverging cold gas evolution seen in the CR and hot wind simulations. We attribute this to differences in our setup and the impact of CRs. First, the cold and hot gas in our simulations do not always experience laminar shearing due to CR streaming along the perturbed magnetic field. Second, the CR heating can be an important external energy source. Although CR heating is relatively small compared to net cooling in mixing gas, it can be important in hot gas. The CR heating in cold gas is also strong but quickly radiated away. These effects change the gas properties and emission in the hot and cold gas that supplies mixing. Finally, the gas pressure in cold and hot gas is not always in balance due to CR and magnetic pressure. These various nonthermal factors complicate quantitative estimation. Dynamically, given the small clouds generated by differential CR acceleration, accounting for the cold gas loss in small clouds is necessary in our setup.

Fielding et al. (2020) showed that the surface area of the mixing layer gas is another important factor impacting cold gas evolution. The efficient mixing creates a corrugated, fractal interface between the cold and hot gas, which generally increases the surface area and promotes cooling. Although the process might not be resolved in our cloud-scale simulations, understanding how the scaling relations change with the presence of CR is an interesting topic for future work.

Previous studies also suggest the importance of resolution of the Field length to reach convergence (Gressel 2009; Sharma et al. 2010a). Although we did not carry out resolution study, with our choice of parameters, the initial Field length is resolved by $\sim$70 cells in the $x$-direction and $\sim$20 cells in the $y$-direction. Interestingly, Tan et al. (2021) suggested that even if the Field length is under-resolved, the total emission from mixing tends to converge if thermal conduction is present. In lieu of a resolution study, we also ran zoomed-in, cloud-crushing simulations of selected irregular clouds formed in thermal instability with three times higher resolution on each side. The overall dynamics is similar to the simulations reported in the paper, and no significant additional small-scale instability is observed.

*Cloud Survival:* In our simulations, thermal instability creates multiple irregular $\sim 10^4$ K clouds, which are accelerated by CRs or a hot wind in a "wind-tunnel"–like setup. Despite the different cloud morphology and nonidealized background, the CR–cloud interaction and cloud survival are qualitatively consistent with idealized single-cloud CR–cloud-crushing simulations. Wiener et al. (2019) first studied the dynamics of a spherical $\sim 10^4$ K cloud that irradiated by transient incoming CR energy flux in hot $\sim 10^6$ K background gas. Despite the different numerical treatments for CR streaming, we find similar elongated cloud morphology due to CR pressure support. When the incoming CR energy flux and magnetic field are relatively strong, the CR "bottleneck" is the main momentum source driving the cloud.

However, with the same cooling function and a lower constant heating rate, Wiener et al. (2019) showed that the radiative cooling is able to seed cold gas growth, while our simulations show cold gas loss with comparable resolution. One difference is that the dynamics in their simulation are nearly one-dimensional and less turbulent. Even with the strongest CR source that can deform the magnetic field, their cloud is primarily accelerated in the $x$-direction in roughly laminar flow. In our setup, the magnetic field is entangled by thermal instability, and the clouds are accelerated in both the $x$- and $y$-directions. By the time we stop the simulation, the velocity dispersion of cold gas in CR-driven cold gas generally reaches $\sim$50–80 km s$^{-1}$ due to the turbulent motion.

In contrast, Brüggen & Scannapieco (2020) suggested that radiative cooling has a limited effect on cloud dynamics and mass evolution. They studied the dynamics of an initially cold cloud irradiated by a constant CR source. The difference in setup includes that they decouples CR momentum with high-temperature background gas. They also do not include CR heating, so CRs cannot directly heat the gas. In our simulation, CR heating contributes a total energy to the multiphase gas that is comparable to the net cooling, and shapes subsequent dynamics. For example, in B05CR, CR heating increases the hot gas thermal pressure and provides initial acceleration to the cold gas. Brüggen & Scannapieco (2020) also explored a wide range of $P_{CR}/P_{gas}$, magnetic field, and density contrast, overlapping with our simulation parameters. For example, the elongated and fragmented cloud morphology seen in their $\beta = 10, 3$ runs with density contrast of 300,100 seems similar to some stretched, fragmented clouds in B1CR and B05CR with comparable local density contrast and $\beta$. With the perturbed magnetic field and irregular cloud shape, however, we did not observe a significant "two-tail structure" near the cloud boundary. Their work also indicates that the cloud acceleration is relatively insensitive to the magnetic field strength, in agreement with our results.

Bustard & Zweibel (2021) investigated the cloud–CR interaction with plasma-based streaming transport. However, their work shows that magnetic field strength changes cloud acceleration given the transient CR flux from the boundary, especially in fully ionized clouds. As in our setup, CRs stream along the field lines, so the topology of the magnetic field affects the CR pressure distribution. In fully ionized clouds, an intermediate magnetic field allows the field lines to wrap around the cloud, preventing CRs from quickly entering the cloud and eliminating the pressure gradient, which is more favorable for strong acceleration. Interestingly, in our simulations, thermal instability forces the field lines to thread through the cloud instead of wrapping around it, which might reduce the CR pressure gradient compared to isolated clouds in an initially plane-parallel magnetic field. They also find that the ionization fraction shapes the "stairs" of the CR pressure gradient at the cloud interface. Relaxing our assumption of fully ionized streaming in multiphase gas formed from thermal instability would be an interesting topic for future work.

Bustard & Zweibel (2021) also compared the effect of dimensionality and found two-dimensional and three-dimensional results differs from one-dimensional results. Gronke & Oh (2020) studied clouds accelerated by thermal wind and suggested that when radiative cooling is included, the cold gas can increase in three-dimensional clouds but decrease in comparable two-dimensional clouds. Investigating our setup in three dimensions will be a useful goal for future work.

## 5. Conclusion

We studied multiphase outflows formed from thermal instability, comparing CR-driven outflows to outflows driven by a hot wind with an equivalent momentum flux (pressure).

We mainly focus on the streaming limit and vary the initial magnetic field (with $\beta_{\rm init} = 0.5, 2, 8$) to study the CR–gas interaction, cold gas survival, and acceleration. We summarize our conclusions as follows:

(1) Density contrast: Unlike a hot wind that compresses the cold gas during acceleration, streaming allows CRs to penetrate into the cold gas, so that CR pressure provides nonthermal support, yielding a lower density for cold gas in the multiphase outflow. In our simulations, the density contrast between $\sim 10^4$ K and $\sim 10^6$ K gas (Figure 6) in CR-driven outflow is $\rho_c/\rho_h \sim 100$, which is lower than the gas density contrast in hot-wind-driven outflow (Section 3.3).

(2) Cold gas evolution: In our simulations, CR-driven outflow shows cold gas loss, while their hot wind counterpart shows cold gas growth. We note that the dropping cold gas, however, does not indicate inefficient cooling. The simulation runs $\sim 20$–$30 t_{\rm cc}$ (Sections 3.2, 3.4), and the cold gas drop is $\lesssim 20\%$ for streaming runs, suggesting that cooling prevents the quick disruption of cold gas. CR streaming does not lead to strong shear between the cold and hot gas in bulk acceleration, so that the different phases are moving with only moderate relative velocity. The cold gas evolution is impacted by both the supply of cold gas from cooling and the dynamical processes destroying cold gas. The origin of cold gas loss varies with magnetic field strength and CR transport mechanism.

When magnetic pressure is subdominant ($\beta_{\rm init} = 2, 8$), the intermediate-temperature gas in the mixing layer is primarily supported by CR (gas) pressure when accelerated by CR streaming (hot wind). CR pressure provides nonthermal support and yields less compression, so the intermediate-temperature gas density is lower. Since cooling is proportional to $n_H^2$, it is reduced compared to the hot wind accelerated outflow where the intermediate-temperature gas is more compressed.

When the magnetic field is strong ($\beta_{\rm init} = 0.5$), intermediate-temperature gas is supported by magnetic pressure, which is similar for both CR and hot wind acceleration. So intermediate-temperature gas density and cooling are similar. But the strong magnetic field constrains the flow motion to be less turbulent. CR streaming forms a "bottleneck" that preferentially accelerates the side of the cloud opposite to the CR source. This leads to the differential acceleration on this side of the cloud, so the fast-moving gas detaches from the bulk and forms small clouds (Section 4.2). We estimate that $\sim 25\%$ of cold gas mass is in the form of small clouds, and nearly half of them are unlikely to survive during acceleration, which largely contributes to the dropping cold gas mass in CR acceleration.

In the slow diffusion limit, CRs compress both hot and cold gas, resulting in overall higher gas densities with $\rho_c/\rho_h \sim 70$, and consequently more cooling. Changing CR transfer from streaming to diffusion alters the evolution of thermal instability, resulting in significant filaments along the magnetic field. The filaments are fragile to even moderate differential acceleration and tend to break into small clouds (Figure 16). We estimate $\sim 20\%$–$40\%$ of cold gas mass is in the form of small clouds, leading to the largest cold gas loss ($\sim 30\%$ by $t' = 70$) among CR runs.

(3) CR streaming in nonuniform magnetic field: B05CR provides an interesting case study of CR streaming in a turbulent magnetic field. First, the initially weak and uniform magnetic field is modified by thermal instability and CR streaming. The field lines thread through the cold gas, and $\beta$ in the $\sim 10^4$ K gas can be an order of magnitude larger than the initial value (Section 4.1).

CRs stream along the perturbed field lines in both the $x$- and $y$-directions, leading to significant CR pressure gradients in both directions. Due to the enhanced magnetic field and CR pressure gradient in the $y$-direction, CR heating in $\sim 10^6$ K gas becomes significant. The thermal pressure of hot gas largely contributes to cold gas acceleration in the $x$-direction. We estimate that the fractional contribution from heated background gas pressure is about five times that of CR pressure (Section 3.5). We find that even in the turbulent, nonuniform magnetic field, CR streaming forms "bottlenecks" along the magnetic field line (Figure 17). Consequently, the turbulent magnetic field is not efficient in transporting CRs via streaming along a preferred direction, resulting in relatively small CR acceleration in the $x$-direction. At the same time, streaming along entangled field lines leads to high velocity dispersion in cold gas ($\sim 60$ km s$^{-1}$).

(4) CR acceleration: In all simulations, we inject a CR flux of $F_{\rm CR} \approx 2.5 \times 10^5 L_\odot$ kpc$^{-2}$ in the $x$-direction. When streaming is enabled, the CR acceleration is relatively linear (Figure 4), as expected from Equation (10). Stronger magnetic fields can slightly increase the acceleration by constraining the flow motion to be more laminar. By the time we stop the simulation ($\sim 100$ Myr), the cold gas ($T \sim 10^4$ K) reaches a mass-weighted average $x$-direction velocity of $v_{\rm cold,x} \approx 210$, 230, and 130 km s$^{-1}$ in B1CR, B2CR, and B05CR, respectively. The velocity dispersion of cold gas is $\sigma_{\rm cold} \approx 35$, 45, and 60 km s$^{-1}$ in B1CR, B2CR, and B05CR, respectively. The average cold gas velocity is similar in their hot wind counterparts. The hot gas $T \sim 10^6$ K is also *accelerated* to nearly comoving with cold gas, with slightly lower average velocity $v_{\rm hot,x} \approx 170$, 180 km s$^{-1}$ in B1CR and B2CR. For B05CR, the turbulent motion accelerates hot gas to $v_{\rm hot,x} \approx 230$ km s$^{-1}$, which is about twice as fast as the average cold gas velocity.

We would like to thank S. Peng Oh, TSun Hin Navin Tsung, Drummond Fielding, and Max Gronke for stimulating discussions and helpful insights. We also appreciate the anonymous referees for the helpful input and constructive suggestions. This work used the computational resources provided by the Flation Institute and the Advanced Research Computing Services (ARCS) at the University of Virginia. We also used the Extreme Science and Engineering Discovery Environment (XSEDE), which is supported by National Science Foundation (NSF) grant No. ACI-1053575. This work was supported by the NSF under grant AST-1616171. X.H. was supported by the Simons Foundation through the Flatiron Institute's Pre-Doctoral Research Fellowship program. The Center for Computational Astrophysics at the Flatiron Institute is supported by the Simons Foundation.

## ORCID iDs

Xiaoshan Huang (黄小珊)  https://orcid.org/0000-0003-2868-489X
Yan-fei Jiang (姜燕飞)  https://orcid.org/0000-0002-2624-3399
Shane W. Davis  https://orcid.org/0000-0001-7488-4468